\documentclass[11pt,letter]{article}
\usepackage{jheppub}

\usepackage{epstopdf}
\usepackage{graphicx}
\usepackage{float}
\usepackage{subcaption}
\usepackage{color}
\usepackage{enumerate}
\usepackage{amsmath}
\usepackage{amssymb}
\usepackage{multirow}
\usepackage{hyperref}
\usepackage{appendix}

\newcommand{\overbar}[1]{$\overline{\text{#1}}$}
\newcommand{\Lim}[1]{\raisebox{0.5ex}{\scalebox{0.8}{$\displaystyle \lim_{#1}\;$}}}
\newcommand{\nl}{\par\vspace{\baselineskip}}

\newcommand{\squ}{u}
\newcommand{\cA}{\mathcal{A}}
\long\def\/*#1*/{}

\title{Infrared Dynamics of a Large $N$ QCD Model, the  Massless String Sector and Mesonic Spectra}
\author{Keshav Dasgupta${}^1$, Charles Gale${}^1$, Mohammed Mia${}^2$, Michael Richard${}^1$, Olivier Trottier${}^1$}

\affiliation{${}^1$Physics Department, McGill University,
3600 University St, Montr{\'e}al, QC H3A 2T8, Canada}

\affiliation{${}^2$Department of Physics, Purdue University,
525 Northwestern Avenue, West Lafayette, IN 47907-2036, USA}

\emailAdd{keshav@hep.physics.mcgill.ca,  gale@physics.mcgill.ca}
\emailAdd{~~~~~~ michael.richard, olivier.trottier@mail.mcgill.ca}
\emailAdd{~~~~~~ mmia@purdue.edu}

\abstract{
A consistency check for any UV complete model for large $N$ QCD should be, among other things, the existence of a well-defined vector and scalar mesonic spectra. In this paper, we use our UV complete model in type IIB string theory to study the IR dynamics and use this to predict the mesonic spectra in the dual type IIA side. The advantage of this approach is two-fold: not only will this justify the consistency of the supergravity approach, but it will also give us a way to compare the IR spectra and the model with the ones proposed earlier by Sakai and Sugimoto. Interestingly, the spectra coming from the massless stringy sector are independent of the UV physics, although the massive string sector may pose certain subtleties regarding the UV 
contributions as well as the mappings to actual QCD.
Additionally, we find that a component of the string landscape enters the picture: there are points in the landscape where the spectra can be improved somewhat over the existing results in the literature. These points in the landscape in-turn also determine certain background supergravity components and fix various pathologies that eventually lead to a consistent low energy description of the theory.}

\begin{document}

\maketitle

\section{Introduction}

Two of the most striking {features} of QCD which can be {captured} in the large $N$ limit are the existence of the confinement to deconfinement transition and the mesonic spectrum, the latter of which is the subject of the present paper. The confining property of QCD, where at low energies the flux lines between quarks become tubular, survives the large $N$ limit even if we remove the whole non-planar sector of the theory \cite{witteN}. 
Thus, both the planar and the non-planar diagrams when summed up individually should provide IR confinement. At high temperatures, the flux tube is broken, whose onset starts at a specific 
temperature $T_c$ called the deconfinement temperature. Beyond $T_c$, the flux lines between the quarks {have} a Coulombic behavior.
Again, in the large $N$ (or planar) limit, this behavior survives: the confinement to deconfinement 
transition\footnote{Note that $SU(3)$ QCD with light flavors has crossover transition from
confining to deconfining. Our understanding is there is still a confining
phase in the non-planar limit. The exact phase transition from confining to
deconfining is realized in the planar limit.}
does happen at a specific temperature scale $T_c$. 

Why is this the case? An answer to this question came surprisingly from a different {field}: black-hole physics, which was hitherto thought to be completely unrelated to the dynamics of QCD. {The first connection between these two theories was established just after the development of gauge/gravity dualities}. The original work {of Maldacena} \cite{maldacena} {described} a duality between certain scale-invariant theories and AdS spaces. In a more modern avatar, the duality can be extended {to link} any gauge theory (with scale dependent couplings) {with a} string theory on a specific background geometry. For small gauge theory couplings, or along the RG trajectory where couplings don't diverge, this duality is not very useful because the string theory side is highly non-trivial. 
Simplification happens in the large $N$ limit: when the 'tHooft coupling of the gauge theory is kept large, {one only needs to analyze the massless sector of the string theory to understand the dual gauge theory.}  

In retrospect, the existence of gauge/gravity duality, although hard to derive analytically, could be argued more intuitively. In the four-dimensional gauge-theory to five-dimensional gravity duality,
the radial direction in the gravity dual is related to the energy scale of the theory. This implies the following statement: moving from one $3+1$ dimensional slice of spacetime at $r = r_1$ to another $3+1$ dimensional slice at $r = r_1 + \delta r$ is equivalent to moving from a $3+1$ dimensional description of a gauge-theory at a scale $E_1$ to another one at a scale $E_1 + \delta E$. This then leads to an intriguing possibility: {Could it be that a five-dimensional gravitational theory is constructed by {\it stacking up} all its dual gauge theory descriptions from far IR to far UV?} This definitely is an interesting way to intuitively 
appreciate the mysterious gauge/gravity duality. It works well in the case of CFTs where, due to scale invariance, we can restrict the gravity dynamics to the AdS boundary. By construction, this meshes well with non-conformal gauge theories too where physics at various scales matters. In the large $N$ limit, the gauge theory planar diagrams are exactly like the string diagrams\footnote{The non-planar diagrams are also string diagrams \cite{tHooft, witteN}.} as shown by 'tHooft \cite{tHooft}, which could 
probably explain how the full duality to string theory can come about. For us, the duality that we are interested in is one that relates a UV conformal to an IR confining gauge theory that mimics QCD, and is given by the geometry first proposed in \cite{FEP}. The IR dynamics is then captured by a warped resolved-deformed cone (which will be 
elaborated in section \ref{secUV}).

It is no surprise that this also works well with the high temperature description of the theory. The high temperature dynamics should be given by inserting a black-hole in the ambient 
geometry, modulo one subtlety. If we denote the geometry without a black-hole by $X^1$ and the geometry with a black-hole by $X^2$, with corresponding on-shell actions ${\cal S}^1$ and ${\cal S}^2$ respectively, then for 
$T>T_c$, black hole geometry $X^2$ will be preferred and we can identify the entropy of the gauge theory with that of the black hole. For $T < T_c$, it is the geometry $X^1$ without a black-hole that is preferred. 
If $T_c$ is large, we can consider black holes in the large $r$ regime and ignore the deformation and resolution.
Then, an exact computation of
$\delta {\cal S} \equiv {\cal S}^2 - {\cal S}^1$ up to linear order in $\delta \equiv g_sM^2/N$ and $g_sN_f$, where the set
($N, M, N_f$) is the number of $D3$, $D5$ and $D7$ branes respectively that we insert in the geometry, is possible. However, the resulting trace anomaly
\begin{equation}
\triangle\equiv \frac{e-3p}{T^4}
\end{equation}
is constant and unlike the one we expect for QCD \cite{recentpapers2}. The reason behind this discrepancy is the underlying  assumption that $T_c$ is large, which allowed us to completely ignore the resolved-deformed region. On the other hand, it is the resolution and deformation of the conifold that give rise to confinement. Thus, we expect $T_c$ to be sensitive to the resolved-deformed region and hence it cannot be large.    

For small $T_c$, we have to consider black holes in resolved-deformed conifold
geometry. However, our perturbative analysis breaks down since $\delta$ is no longer a small quantity near the tip of the deformed cone. Thus, we cannot evaluate $\delta {\cal S}$ perturbatively and cannot determine $T_c$.  Nevertheless, we can find the scaling of the entropy $s$ with the horizon radius using, {with the additional approximation of} zero resolution at $r = 0$, the form of the deformed cone metric in  Wald's formula \cite{recentpapers2}:
\begin{equation} \label{entropyrho}
s\sim \sqrt{h(\rho_h)}{\rm sinh}(\rho_h)\left({\rm sinh}(2\rho_h)-2\rho_h\right)^{1/3}
\end{equation}
where $h(\rho)$ is the warp factor as a function of the radial coordinate $\rho$, whose details will be specified soon\footnote{The actual radial coordinate $r$ is related to $\rho$ via $r = r_0 e^\rho$. See also \eqref{redef}. In this 
language, $\rho_h$ will be the horizon radius.}.
Since we consider {the} warp factor $h(\rho)$ to be regular near {the} horizon, we get $T(\rho_h)=t_i\rho_h^i$ where $t_i$ are constants. For {a} choice of {the metric's} Taylor coefficients near the horizon, we can numerically solve for the black hole factor and obtain
the coefficients $t_i$. Then, using Wald's formula, we obtain the entropy as a function of temperature and the following scaling for
the free energy \cite{recentpapers2}:
\begin{eqnarray} \label{F-R1}
F&\sim& -T^2  ~~~~ {\rm for}~~T_c < T < {\rm small}~~ T_0\nonumber\\
&\sim&-T^3~~~~
{\rm for}~~ T_c < T < {\rm intermediate} ~~T_0\nonumber\\
&\sim&-T^4\left[1+\frac{b_{01} g_sM^2}{4N}
+\frac{b_1 g_sM^2}{N}{\rm log}\left(T\sqrt{N\alpha'}\right)\right]~~~~ {\rm for}~~T\lesssim {\rm large}~~T_0\nonumber\\
&\sim & -T^4~~~ {\rm for}~~T> {\rm large}~~T_0
\end{eqnarray}
where $T_0$ is determined by the scale $r_0$, $b_{01} \equiv \left(b_0 - \frac{b_1}{4}\right)$ and ($b_0,b_1$) are constants that arise 
from the form of the metric at large radial
distance \cite{recentpapers2}. 
Finally, using these scalings, we get the conformal anomaly for the dual gauge theory \cite{recentpapers2}:
\begin{eqnarray}\label{Anomaly-R1}
\triangle&\sim& \frac{1}{T^2} ~~~{\rm for}~~ T_c < T < {\rm small} ~~T_0 \nonumber\\
&\sim& \frac{1}{T}~~~~~{\rm for}~~ T_c < T < {\rm intermediate}~~ T_0 \nonumber\\
&\sim& \frac{27\pi^6  N\alpha'^4V_5b_1 g_sM^2}{512\kappa_{10}^2}~~{\rm for}~~T\lesssim {\rm large}~~T_0\nonumber\\
&\sim& 0~~~~~~~{\rm for}~~T> {\rm large}~~T_0
\end{eqnarray} 
where $V_5$ is the volume of the internal space. 
Note that the scaling of the anomaly is similar to {the one derived from} lattice simulations \cite{Panero:2009tv}. 

The above discussion on the thermodynamics of large $N$ QCD is encouraging and demonstrates the power of holography in studying the non-perturbative regimes of QCD. One {would then hope for a} similar success in understanding the mesonic spectra {of} large $N$ QCD. {In this case, holography faces stronger constraints} because we expect the gravity description to reproduce the {\it linear} Regge trajectories:
\begin{equation}
J = \alpha_0 + \alpha' M^2
\end{equation}
where $J$ is the spin, $M$ is the mass of the Hadrons, $\alpha_0$ is a constant and $\alpha'$ is the Regge slope. The original attempt of string theory to reproduce this property failed as the string spectra contained massless spin 2 states. {Hence, string theory was proclaimed} a theory of gravity, {while} QCD {remained} the unique theory of hadronic interactions. {At the advent of gauge/gravity dualities, string theory reclaimed its fame, but not for the reasons originally envisioned. Providing a {\it dual} description of large $N$ QCD with gauge/gravity dualities, string theory was reconsidered as a candidate theory of hadronic interactions. }

The first solid attempt to reproduce the mesonic spectra from the gravity dual was performed by Sakai-Sugimoto \cite{SS, SS2} using type IIA string theory. This work is remarkable {for many reasons}: not only {did it provide} a working model for large $N$ QCD, but {it} also gave a way to reproduce the 
spectra using open string dynamics in the dual side, albeit using AdS/CFT. 
However, it didn't quite reproduce the Regge behavior as the analysis of \cite{SS, SS2} were restricted to the massive Kaluza-Klein (KK) states in the massless open string sector. The analysis was improved more recently in \cite{SSrecent}. {The work of Sakai and Sugimoto demonstrated yet another application of gauge/gravity dualities, i.e., the ability to study the mesonic spectra from a gravitational point of view. Moreover, it motivated others to study the mesonic spectra with other string theories \cite{other0, other1, other2}, for which the type IIB Klebanov-Strassler model \cite{KS} is an example.} However, {no one so far has undertaken the analysis of} the spectra using a UV complete model, like \cite{FEP, uvcomp}. Furthermore, a {\it direct} comparison with Sakai-Sugimoto model on the type IIA side using the T-dual of the UV complete model has never been {tried before}. Our work attempts to address these two challenges.

\subsection{Organization of the paper}

The paper is organized as follows. In section \ref{secUV}, we discuss in some details the UV complete model first proposed in \cite{FEP} and then delve deeper into the issues of stability in section \ref{secBG}. The T-dual version of the type IIB setup is then elaborated in section \ref{secIIA} followed by the D6-brane embeddings in section \ref{secD6}. {In section \ref{SS}, we describe the {type} IIA dual and the D6-brane embeddings, which are necessary to compare our model with the one proposed by Sakai-Sugimoto \cite{SS}}. Section \ref{secvec} {introduces} our main computations of the vector mesonic action, {followed} with {the} details of the spectrum {calculations} in section \ref{S:VM}. The analysis of the vector mesonic spectra depends on our choice of the expansion parameter $\delta$. {Section \ref{seczero} \& \ref{secfirst} present our zeroth and first-order approximations of the set of eigenvalues that are used to calculate our mass-ratio predictions.} The results of the vector mesonic spectra are {summarized} in section \ref{secmeson}.   

In section \ref{UVphysics}, we take a short detour to analyze the effects of the UV physics on the spectra. We discuss how the UV regimes in the supergravity dual may not affect the IR physics, at least for the massless open string sector of the theory.  

{Mimicking the techniques used for the vector mesonic spectrum, we derive the scalar mesonic action in section \ref{secmesac} and describe the spectrum in section \ref{secmess}. More precisely, {the} zeroth and first-order computations are presented in sections \ref{secmeso} and \ref{secmes1} respectively. Although the scalar and vector mesonic spectrum calculations share many similarities, the identification of their particle content with well-known QCD particles posed different difficulties: there is a certain $Z_2$ symmetry expected for QCD scalar mesons that is not shared by the scalar mesons got from the massless open string sector. Thus, although the scalar mesonic spectrum is recovered from the open string sector and may not be related to actual QCD particles, the vector mesonic spectrum could still be compared, modulo the subtlety the massive open string states may pose. Since the identification with real QCD particles could not be established, we omitted our scalar mesonic predictions in our comparison with other holographic QCD models. Finally, section \ref{secmesid} assembles all our knowledge of the scalar mesonic spectrum. 

Dependence of the spectra on the $\theta_2$ expansion is then discussed in section \ref{sec6}. We also find that there are other values of the expansion parameter $\delta$ and {the squashing factor} $\squ$ for which the spectrum can be improved somewhat. This is where the landscape of flux vacua enters and we elaborate this in section \ref{landscape}. We end with comments on our findings in section \ref{seccono} and appendices \ref{land} and 
\ref{landb} {describe our predictions at} other points in the landscape of flux vacua including its boundary.

\section{The UV complete large $N$ QCD model \label{secUV}}

The story begins in \cite{FEP} where the first dual gravitational description for a specific ultra-violet (UV) complete large $N$ QCD model was proposed\footnote{The model is not quite SQCD as the issue of supersymmetry is a little subtle here, and we will discuss this later. We will also discuss the stability of our model soon.}. Soon after, a detailed analysis of the UV region and the consequent dual geometry were presented in \cite{uvcomp}. Our analysis showed that the gravity dual can be succinctly divided into three regions. Region 1 captures the deep infra-red (IR) description wherein chiral symmetry breaking, gluino condensation and linear confinement can be understood. {Region 3 captures the dynamics of the UV region where the gauge theory approaches asymptotic conformality. Finally, Region 2 is the intermediate region where the theory is interpolating from a confining IR dynamics to a conformal UV}. 

On the gauge theory side, the dynamics is equally easy to understand. In the far UV, the theory approaches conformality and the color gauge group is 
$SU(N+M) \times SU(N+M)$. The walking RG flows for the two couplings $g_1$ and $g_2$ of the two gauge groups {arise} from the fundamental matter multiplets. The color factors $N$ and $M$ are related, on the gauge theory side, to the $N$ D3 and $M$ D5-branes. The fundamental matter multiplets appear from $N_f$ D7 and anti-D7 pairs, giving rise to the $SU(N_f) \times SU(N_f)$ global symmetry. 

At an intermediate scale, one of the $SU(N+M)$ gauge groups is Higgsed \cite{higgsing}, changing the color group to $SU(N+M) \times SU(N)$. {At this point,} the two couplings $g_1$ and $g_2$ flow differently. This is where the scenario becomes more subtle because {a difference in the} RG flow means that one of the two couplings is bound to become weak, while supergravity with small curvature at every $r$ describes {only} large 'tHooft couplings.

To clarify the situation a bit more, note that the dual gravitational description that we seek should not have the stringy corrections incorporated in. This is only possible if the 'tHooft coupling of the gauge theory remains strongly coupled from UV to IR. {Using} the AdS/CFT correspondence, this is achieved by allowing the CFT coupling to be infinitely large. This is {clearly} possible for our UV regime as we can keep both $g^2_1(N+M)$ and $g^2_2(N+M)$ infinitely large (even with the walking RG flow). 

At an intermediate energy range, when the two couplings flow in opposite directions via the NSVZ beta function, the gauge theory will run into a regime where one of the 'tHooft coupling {is no longer large}. In fact, under a Seiberg duality, a gauge theory with large coupling will map to a gauge theory with small coupling and, under a cascade of Seiberg dualities, the far IR gauge group becomes $SU(M)$. The IR dynamics of $SU(M)$ theory is strongly coupled and confining, and therefore all DOFs {are eventually confined}. Since the supergravity description corresponds to a large 'tHooft coupling, the confining regime of the gauge theory is indeed dual to the small $r$ region with classical description\footnote{The small $r$ region is the warped resolved-deformed region, which again has a large curvature radius. When the resolution vanishes, the warped deformed cone geometry is described in \cite{KS}.}. However, for intermediate $r$, supergravity will again describe large curvature radius dual to large 'tHooft coupling and hence cannot fully capture the Seiberg dualities. Thus, for intermediate energies, there is no exact classical gravitational description of the cascade. Nevertheless, classical gravity will capture some features of cascading theories and, as long as we are only interested in the far IR and far UV dynamics, the dual geometry description will be exact.

\subsection{A brief detour into the model \label{secdetour}}

The IR picture, originally developed in \cite{KS}, was modified to include D7 branes in \cite{ouyang, ouyang2}. It was further extended to include black holes \cite{FEP} to have a thermal QCD theory. In \cite{FEP}, we called it the "OKS-BH model" and it is also the Region 1 of our model. 

To avoid certain logarithmic divergences in the KS model, UV completion became necessary.  
Therefore, we modified the model by adding the UV completion to have a consistent theory of QCD. The new theory replaces OKS-BH model at radius $r=r_{min}$. This is where Region 2 begins, which is at the same energy scale as the mass of the lightest quark. Region 2 is an intermediate region that leads to a smooth transition from OKS-BH model to the UV completed geometry. In region 2, the NS three-form gradually decays away. When we step into Region 3, there are no more three-forms, but there is still {a non-vanishing} axio-dilaton (defined as $\tau=C_0 + i e^{-{\phi}}$). This is of course the asymptotic AdS cap that we discussed above.

The theory becomes conformal as $r \rightarrow \infty$ as the RG flow leads to constant couplings. Also, the Wilson loop for heavy quark and anti-quark pairs has no divergence at large radius. D7 and anti-D7 branes, which act as sources of fundamental flavours, spread out from Region 3 to 2, but in Region 1, there is only one coincident set of D7 branes. This breaks the global symmetry from $SU(N_f) \times SU(N_f)$ to $SU(N_f)$. The tachyons between the branes and the anti-branes are cancelled by world-volume fluxes. 

On the gauge theory side, the configuration is given in {\bf figure \ref{condet}}. The details of this configuration has already appeared in our earlier papers \cite{recentpapers} so we will be brief. It is easy to see that at high energy, i.e the energy bigger than the excitation energy of the D5 anti-D5 strings, the theory will be governed by an $SU(N+M_\epsilon) \times SU(N+M)$ gauge theory where the factor $M_\epsilon$ appears from the M D5 anti-D5 pairs (there are no tachyons, so only the D5 charges cancel giving rise to M fractional D3-branes) as:
\begin{equation}\label{Mdist}
M_\epsilon = {M e^{\alpha(r-r_3)}\over 1 + e^{\alpha(r-r_3)}}
\end{equation}
where $r_3$ is the boundary between Region 3 and Region 2, $\alpha >> 1$ and $r \sim {\cal O}(1/\epsilon)$.
At low energies, we may integrate out the anti-D5 brane DOFs and so the theory is $SU(N+M) \times SU(N) \times U(1)^M$,
where the $U(1)^M$ appear from the massless sectors of the $M$
distributed anti-D5 branes and are basically decoupled (the 
D5 anti-D5 strings are integrated out). This is then
similar to the Ouyang-Klebanov-Strassler model. 

Regarding the stability of the model, the reason is simple to state. First, the tachyons between D5 and the anti-D5 branes are made massive by the world-volume fluxes. Therefore, the anti-branes could be placed at any relative distances from the branes without the possibility of an annihilation. Second, to simplify the computation, all forces between the branes can be made independent of the angular directions of the resolved sphere so as to depend only on the radial direction $r$. Thus, both the attractive gravitational force and the {\it repulsive}\footnote{The world-volume fluxes on the branes and the anti-branes effectively create bound D3-branes respectively on the branes and the anti-branes. These set of bound D3-branes in-turn repel each other.} RR force between the D5 and the anti-D5 branes are cancelled out leading to stability. The interesting thing is that even when the background supersymmetry is broken (by our choice of the embedding or by a black-hole insertion or by both), one should be able to choose appropriate world-volume fluxes respectively on the branes and the anti-branes to restore stability. 

Although the above analysis of stability is intuitively clear: the gravitational and the D5-anti-D5 attractions being balanced by the bound branes' repulsion\footnote{There are three kinds of repulsive forces in action here: forces between the $M$ wrapped D5-branes, forces between the $M$ distributed anti-D5 branes and forces between the $M$ D5 and the $M$ anti-D5 branes.}, an exact quantitative analysis is hard to perform. This is because an exact computation will require studying the quantization of strings on a curved background to infer how the tachyons become massive. Despite this analytical hurdle, it should be clear that stability could be achieved for our case by balancing {these} two kinds of forces.

Similar arguments extend to the flavor D7 and the anti-D7 branes also. Killing the tachyons between them using world-volume fluxes would also restore stability to the system. What about forces between the five branes and the seven branes? Since the seven branes are treated as 
probes ($g_sN_f << 1$), they do not alter the supergravity background. Thus, they also do not create any forces between the branes. In general however, the anti-D5 branes could form bound states with the D7-anti-D7 pairs and the system can be made stable {with the} world-volume fluxes as discussed above. But in the probe limit, these details will not {affect} the background. 

\subsection{Background supergravity solution \label{secBG}}

Our discussion above should clarify how a stable but non-supersymmetric UV complete gauge theory could be constructed using type IIB branes, fluxes and anti-branes. The gravitational dual description converts the $N$ D3 and the $M$ wrapped D5 branes into geometry. Anything else in the brane picture (see {\bf figure \ref{condet}}) would {\it remain} in the geometry, but now distributed along the radial direction $r$. This in particular means that the anti-D5 branes on the brane side will continue as anti-D5 branes in the gravity dual and will now be distributed continuously from $r = r_{min}$ to $r = \infty$, i.e from Region 2 onwards to asymptotic infinity as in \eqref{Mdist}. Similarly, the D7 and the anti-D7 branes in the brane configuration will continue to be D7 and anti-D7 branes in the dual gravity description. The energy scale in the brane configuration will now be the radius scale of the dual theory. 
Once the dual background is constructed, it is not too hard to write the full supergravity solution. This has been worked out in details in \cite{uvcomp, recentpapers, recentpapers2}. For our purpose, it will suffice to concentrate only on Region 1. The background metric for the dual strongly coupled theory in this region is given by:
\begin{equation}\label{bhmat}
ds^2 = {1\over \sqrt{h}}
\Big[-\text{dt}^2+\text{dx}^2+\text{dy}^2+\text{dz}^2\Big] +\sqrt{h}\Big[dr^2+ r^2 d{\cal M}_5^2\Big]
\end{equation}
where $h(r)$ is the warp factor and $d{\cal M}_5^2$ is the metric of the internal non-compact space and whose value is given by:
\begin{eqnarray}\label{sukra}
d{\cal M}_5^2  & = & h_1(d\psi + {\rm cos}~\theta_1 d\phi_1 + {\rm cos}~\theta_2 d\phi_2)^2 + h_2 (d\theta_1^2 + {\rm sin}^2 \theta_1 ~d\phi_1^2) \nonumber\\
&& + h_3(d\theta_2^2 + h_4 {\rm sin}^2 \theta_2 ~d\phi_2^2) + h_5~{\rm cos}~\psi \left(d\theta_1 d\theta_2 - 
{\rm sin}~\theta_1 {\rm sin}~\theta_2 d\phi_1 d\phi_2\right) \nonumber\\
&& ~~~~~~~~~~~~~~~~~~~~~~~~ + h_5 ~{\rm sin}~\psi \left({\rm sin}~\theta_1~d\theta_2 d\phi_1 - 
{\rm sin}~\theta_2~d\theta_1 d\phi_2\right)
\end{eqnarray}

\begin{figure}[H]
\centering
\includegraphics[width=\textwidth]{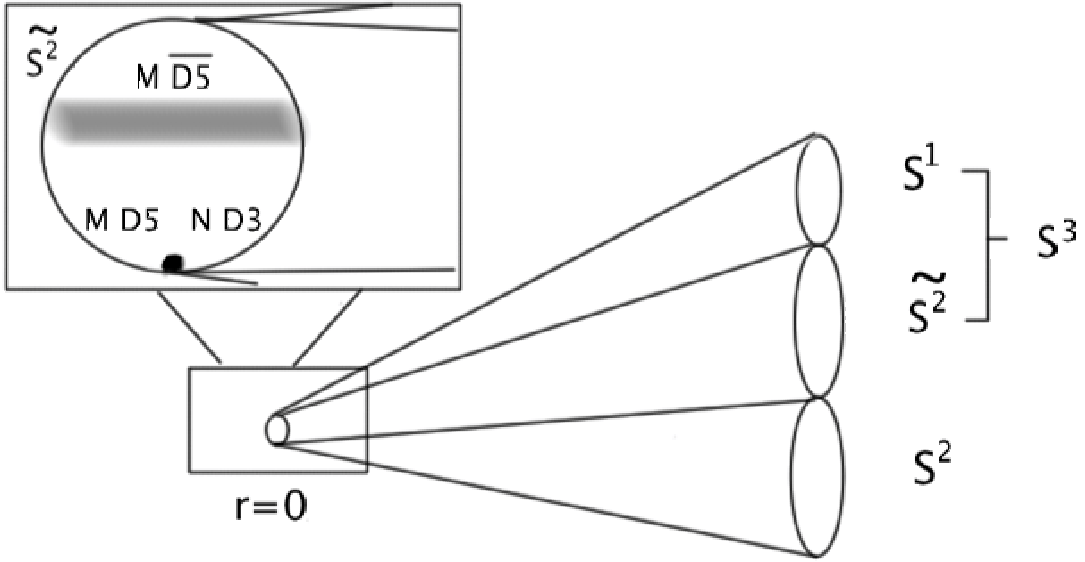}
\caption{Brane construction for the gauge theory and details of the non-K\"ahler resolved conifold geometry.}
\label{condet}
\end{figure}

The $h_i$ denoted above are the internal warp factors, and note that one of the two-spheres is squashed by $h_4$. A derivation of this squashing has appeared in \cite{beckers} where one may also get the values for $h_i$.
For small $r$ and in the limit where we consider the D7-branes as probes, the values for $h_i$ are not too hard to find. They are given by:
\begin{equation}\label{squaf}
h_1 = 1, ~~~ h_2 ~ = ~h_3 ~ = ~ {1\over 6}, ~~~ h_4 = 1 + {u\over {\rm sin}^2\theta_2}, ~~~ h_5 ~ = ~ u_0
\end{equation}
where ($u, u_0$) are two fixed quantities whose explicit values will be discussed soon. Additionally, the axionic field from the D7 and anti-D7 branes will cancel in the probe limit. Away from the probe limit, the axionic field will be proportional to ${\cal O}(g_sN_f)$ for $N_f$ seven-branes. The dilaton will be constant and the five-form field strength will take the standard form given in terms of the warp factor $h(r)$. On the other hand, the three-form field strengths take the following values:
\begin{eqnarray}\label{mrkoremi}
{\widetilde F}_3 & = & 2M  {\bf A_1}~e_\psi \wedge 
\frac{1}{2}\left({\rm sin}~\theta_1~ d\theta_1 \wedge d\phi_1-{\bf B_1}~{\rm sin}~\theta_2~ d\theta_2 \wedge
d\phi_2\right)\nonumber\\
H_3 &=&  6g_s M {\bf A_4}~ \frac{dr}{2r}
\wedge \left({\rm sin}~\theta_1~ d\theta_1 \wedge d\phi_1
- {\bf B_4}~{\rm sin}~\theta_2~ d\theta_2 \wedge d\phi_2\right)
\end{eqnarray}
where $\widetilde F_3 \equiv F_3 - C_0 H_3$, $C_0$ being the ten dimensional axion and
the so-called asymmetry factors ${\bf A_i}, {\bf B_i}$ are given in eq. (3.83) of \cite{FEP}.

One may now similarly work out the supergravity backgrounds for Regions 2 and 3. These have been studied earlier, so we will not discuss them here. {Instead, we will study the brane configurations and the dual gravitational background of the alternative type IIA description.} This approach will allow us to compare our model to the type IIA Sakai-Sugimoto model.

The type IIA description should be achieved by performing a T-duality along the $\psi$ direction. This is more easily said than done. As one looks at the background metric, \eqref{bhmat} and \eqref{sukra} tell us that we have lost
the isometry along the $\psi$ direction because of the deformed conifold structure! One way to regain the isometry will be to impose:
\begin{equation}\label{h5def}
h_5 ~ = ~ u_0 ~ = ~ 0
\end{equation}
but this will give rise to a conifold geometry at $r = 0$ destroying the IR confining behavior on the dual gauge theory side. This is of course {\it not} what we need. A suitable compromise, that will not change the background EOMs but still allow us to make a T-duality, will be to put a {\it cut-off} at $r = r_0$ so that the deformation 3-cycle is hidden behind $r = r_0$. This means, for $r \ge r_0$, the geometry will {look as though it was a cone with its metric given by \eqref{sukra} and \eqref{h5def}.} In other words, we will assume 
\begin{equation}\label{redef}
r ~ = r_0 e^\rho  ~ \equiv ~ {\cal A}^{2/3} e^\rho
\end{equation}
with $\rho$ as the new radial coordinate, and ${\cal A}$ is a constant. On the gauge theory side, this would be like putting an IR cutoff at very low energies.

Before moving ahead, one may question the meaning of the mesonic spectra in a theory with an IR cutoff where the far IR dynamics have been eliminated. There are two ways to answer this question. First, we're only putting a very low energy cut-off (to facilitate subsequent T-duality), so most of the low energy regime is readily available. There is no cut-off in the original type IIB side\footnote{In this sense it may differ from hard wall QCD computations, although it will be interesting to compare the two scenarios.}. Second, the IR dynamics manifests itself via the choice of the cutoff in the following sense: $r_0$ determines a scale and the masses of the IR degrees of freedom, i.e. the mesons are measured in units of $r_0$. Thus, the cutoff carries information about low energies and {its particular value influences the way we probe the} IR physics. Since we fix $r_0={\cal A}^{2/3}$ to reproduce the observed mesonic spectra in QCD, our cutoff is consistent with the IR of QCD.

\section{Background fields in type IIA theory \label{secIIA}}

Having clarified the basic construction of our model, it is now time to go to the type IIA side.
After T-dualizing along the $\psi$ coordinate of the conifold geometry, we obtain the following background metric in type IIA:
\begin{equation}\label{2amet}
ds^2 = \frac{-\text{dt}^2+\text{dx}^2+\text{dy}^2+\text{dz}^2}{\sqrt{h(r)}} + \frac{9 L^4 }{r^2 \sqrt{h(r)}}\text{d$\psi$}^2 + \sqrt{h(r)} \left(\text{dr}^2 + r^2 \, d\Sigma^2\right)
\end{equation}
where $d\Sigma^2$ specifies the metric of the internal space, and is now given by:
\begin{equation}
d\Sigma^2 \equiv \frac{1}{6} \left[\text{d$\theta $}_1^2 + \text{d$\theta $}_2^2 + \sin ^2\theta_1 \text{d$\phi $}_1^2  + \left( \squ + \sin ^2\theta _2 \right) \text{d$\phi $}_2^2 \right]
\end{equation}
The NS $B$-field now has two-sources, one from the original type IIB $B$-field and the other from the 
$d\psi$ fibration structure, and is expressed in the following way:
\begin{align}\label{bilu}
B = &\, 3 g_s M \log (r/\mathcal{A}^{2/3}) \left(\sin ~\theta_1 \text{d$\theta $}_1\wedge \text{d$\phi $}_1 + \sin~\theta _2 \text{d$\theta$}_2\wedge \text{d$\phi $}_2\right)\nonumber\\
& ~~~~~~~~~~~~+ 2 L^2 \text{d$\psi $} \wedge \left( \cos~\theta _1 \text{d$\phi $}_1 + \cos~\theta _2 \text{d$\phi $}_2 \right)
\end{align}
The dilaton is not a constant, but is given by:
\begin{equation}\label{dilla}
e^{-\phi(r)} = \frac{h(r)^{\frac{1}{4}} r}{6 g_s}
\end{equation}
Finally, the warp factor $h(r)$, in the limit where the flavor seven-branes are probes, takes the following form in Region 1:
\begin{equation}
h(r) = \frac{27 \pi g_s N}{4\, r^4}  \left[1 + \frac{3 g_s M^2  \log (r/\mathcal{A}^{2/3})}{2 \pi  N}\right]
\end{equation}
where $\mathcal{A}^{2/3}$ is the minimal value of $r$ and $\squ$ parametrizes the squashing of one of the two conifold  spheres discussed earlier in the type IIB context. Note that both the NS B-field \eqref{bilu} and the dilaton \eqref{dilla} are kept
independent of the squashing parameter $\squ$. This is possible because in type IIB, all the RR fields: $\widetilde{F}_3, F_3, C_0$ and
$\widetilde{F}_5$ {can} be $u$-dependent. This way:
\begin{equation}
H_3 \equiv {\widetilde{F}_3 - F_3\over C_0}
\end{equation}
can be made $u$-independent. Alternatively, we could lift the original type IIB background to M-theory, where the axio-dilaton forms
part of the 11-dimensional metric and ($F_3, H_3$) become the $G$-flux. Thus, the $u$-independence of $H_3$ and dilaton amounts to the $u$-independence of certain components of the $G$-flux and the real part of the complex-structure of the 11-dimensional torus.

At this stage, we will perform the following coordinate transformations, which are more natural to use with our D6-brane embedding.
\begin{align}
&Y = \rho \cos~\theta,& \quad &Z = \rho \sin~\theta \nonumber\\
&\rho = \sqrt{Y^2 + Z^2},&  &\theta = \arctan\left(\frac{Z}{Y}\right) \nonumber\\
& r = \mathcal{A}^{2/3} e^{\rho},& \quad &\psi = c \, \theta \label{EQ:rdef}
\end{align}
$c$ is some constant and $0 \le \rho \le {\rm log}\left({r_{min}\over {\cal A}^{2/3}}\right)$. Since we want to restrict the analysis to Region 1 with $r_0 = {\cal A}^{2/3}$ very small, i.e of ${\cal O}(\epsilon)$ and $r_{min}$ finite and large, $\rho$ and $Z$ have the following range:
\begin{equation}
\rho~\in~\left[0, \infty \right], ~~~~~~~ Z~\in~ \left[-\infty, + \infty\right]
\end{equation}
This means that the dynamics will only happen in Region 1 parametrized by the ($Y, Z$) coordinates, i.e our analysis will capture the IR physics. In the $(Y,Z)$ coordinates, the metric takes the following form:
\begin{align}
ds^2_{(r,\psi)} &=  \frac{9 L^4}{r^2 \sqrt{h(r)}}\text{d$\psi$}^2 + \sqrt{h(r)}\text{dr}^2 \nonumber\\
&= \frac{\left[e^{\sqrt{Y^2 + Z^2}} \left(Y^2+Z^2\right)\right]^{-2}}{{\cal{A}}^{4/3}
   \sqrt{h\left(r(Y,Z)\right)}} \left[A(Y,Z) \left(\text{dY}^2 + \text{dZ}^2 \right) + 2 B(Y,Z) \text{dYdZ} \right]
\end{align}
The coefficients $A$ and $B$ are functions of ($Y, Z$) and are given by:
\begin{align}
 A(Y,Z) &\equiv  {\cal{A}}^{8/3} e^{4 \sqrt{Y^2 + Z^2}} \left(Y^2+Z^2\right) h\left(r(Y,Z)\right)  Y^2  + 9 c^2 L^4 Z^2 \nonumber\\
 B(Y,Z) &\equiv YZ \left[{\cal{A}}^{8/3} e^{4 \sqrt{Y^2 + Z^2}} \left(Y^2+Z^2\right) h\left(r(Y,Z)\right) -9 c^2 L^4 \right]
 \end{align}
The NS $B$-field now can be rewritten in terms of 
($Y, Z$) as:
\begin{align}
B &= 3 g_s M \log \left[{r(Y,Z)\over \mathcal{A}^{2/3}}\right] \left(\sin~\theta _1\text{d$\theta $}_1\wedge \text{d$\phi $}_1 + \sin~\theta _2 \text{d$\theta$}_2\wedge \text{d$\phi $}_2\right)\nonumber \\
  & ~~~~~+ \frac{2 \,c\, L^2 }{Y^2+Z^2} \left( Y \text{dZ} - Z \text{dY} \right) \wedge \left( \cos~\theta _1\text{d$\phi $}_1 + \cos~\theta _2 \text{d$\phi $}_2 \right)
\end{align}

\section{D6-Brane Embedding \label{secD6}}

We proceed by embedding a stack of $N_f$ D6-branes in this background using the probe approximation ($N_f = 1$). We choose the first branch of the Ouyang embedding where $(\theta_1, \phi_1) = (0, 0)$ and we consider the $\psi$ coordinate as a function of $r$, i.e $\psi(r)$. We then use the equations of motion of this field to find the explicit functional dependence. The D6-branes are embedded along $x^{0, 1, 2, 3}$ and wrap the internal three-cycle parametrized by ($\theta_2, \phi_2$) and $\psi(r)$. The pull-back of the metric is given by:
\begin{align}
{g_6}_{MN} dX^M dX^N & = \frac{-\text{dt}^2+\text{dx}^2+\text{dy}^2+\text{dz}^2}{\sqrt{h(r)}} + \left[\frac{9 L^4 }{r^2 \sqrt{h(r)}} {\psi'(r)}^2 + \sqrt{h(r)} \right] \text{dr}^2 \nonumber\\
&~~~~~~~~~~~~~~~~~~~~~~~+ \sqrt{h(r)} r^2 \, \left[\text{d$\theta $}_2^2 + \left(\squ + \sin^2\theta_2 \right) \text{d$\phi $}_2^2 \right] 
\end{align}
whereas the pull-back of the $B$-field along the direction of the D6-branes is given by:
\begin{equation}
B_6 = 3 g_s M \log (r/\mathcal{A}^{2/3}) \sin~\theta _2 \text{d$\theta$}_2\wedge \text{d$\phi $}_2 + 2 L^2 \psi'(r) \cos~\theta _2 \text{dr} \wedge \text{d$\phi $}_2
\end{equation}
Substituting the embedding in the metric and $B$-field presented above, we obtain the following DBI action:
\begin{equation}
\sqrt{-\text{det}(g_6 + B_6)} = 
\sqrt{\frac{81 \left[ g_s M \log\left(\frac{r}{A^{2/3}}\right)\right]^2 \sin ^2\theta_2 {\cal C}_1 + r^4 h(r) {\cal C}_2 +r^2 \psi '(r)^2 {\cal C}_3}{324 L^2 h(r)^{3/2}}}
\end{equation}
where the coefficients ${\cal C}_i$ are given by:
\begin{equation}
{\cal C}_1 = 9L^2 + r^2 \psi'(r)^2, ~~~~~~{\cal C}_2 = 3L^2 (2+3u + \sin^2\theta_2), ~~~~~~ {\cal C}_3 = u + \sin^2\theta_2
\end{equation}

As detailed below (see section \ref{S:VM}), our analysis treats $\delta \equiv \frac{g_s M^2}{N}$ as a small parameter, which we expand up to first order. Therefore, for consistency, we also perform the embedding analysis up to first-order terms in the $\delta$ expansion. First, we derive the Euler-Lagrange equation for $\psi(r)^{(0)}$ from the DBI action.
\begin{equation}
- A \left( (2 + 3\squ+\sin ^2\theta_2) \left(r {\psi''}^{(0)}(r)+5 {\psi'}^{(0)}(r)\right) + 12 r^2 (\squ+\sin ^2\theta_2) {\psi'}^{(0)}(r)^3 \right) = 0
\end{equation}
where the coefficient $A$ takes the following form:
\begin{equation}
A \equiv \left(\frac{\pi^3}{108 g_s^5 N} \right)^{1/4} \frac{r^4 \left(\squ+\sin ^2\theta_2\right)}{3 \left(\squ+\sin ^2\theta_2 + 3 r^2 \left(2 + 3 \squ+\sin ^2\theta_2\right) {\psi'}^{(0)}(r)^2\right)^{3/2}}
\end{equation}
As we can see, $\psi(r)^{(0)} = \text{constant}$ solves the equation. Setting $\psi(r)^{(0)} = \text{constant}$ in the first-order equation, we have:
\begin{align}
 \left(\frac{\pi^3}{108 \, g_s^5 N} \right)^{1/4} \frac{r^4 \sqrt{\squ+\sin ^2\theta_2} \left(4+6 \squ+3 \sin ^2\theta_2-\sin ^4\theta_2\right) \left(5{\psi'}^{(1)}(r)+r {\psi''}^{(1)}(r)\right) }{3 \left(2 \squ+3 \sin ^2\theta_2-\sin ^4\theta_2\right)} =0
\end{align}
Again, we see that ${\psi}^{(1)}(r) = \text{constant}$ is a solution. Since our spectrum analysis does not go further than first order in $\delta$, we may use a constant embedding for the $\psi$ coordinate of the D6-branes. Since the constant is arbitrary, we set the D6-branes and \overbar{D6}-branes at antipodal positions along the $\psi$ direction, respectively. In other words, $\theta = \frac{\pi}{2}$ for the D6-branes and $\theta = -\frac{\pi}{2}$ for the \overbar{D6}-branes. In the $(Y,Z)$ coordinates, this embedding is equivalent to setting $Y = 0$, which considerably simplifies the metric $G$:
\begin{align}
ds^2 &= {g_6}_{ M N} dx^M dx^N \notag\\
&= \frac{-\text{dt}^2+\text{dx}^2+\text{dy}^2+\text{dz}^2}{\sqrt{h(r(0,Z)}} + \sqrt{h(r(0,Z))} {\cal{A}}^{4/3} e^{2 |Z|} \left( \text{dZ}^2 + \text{d$\theta $}_2^2 + \left(\squ + \sin^2\theta_2 \right)\text{d$\phi $}_2^2 \right) 
\end{align}
and the pull-back of the $B$-field along the directions of the D6 and the anti-D6 branes:
\begin{align}
B_6 =  3 g_s M \log \left({r(0,Z)\over \mathcal{A}^{2/3}}\right)\sin~\theta_2 ~\text{d$\theta$}_2\wedge \text{d$\phi $}_2
\end{align}

\vskip.1in

\begin{figure}[H]
\centering
\includegraphics[width=\textwidth]{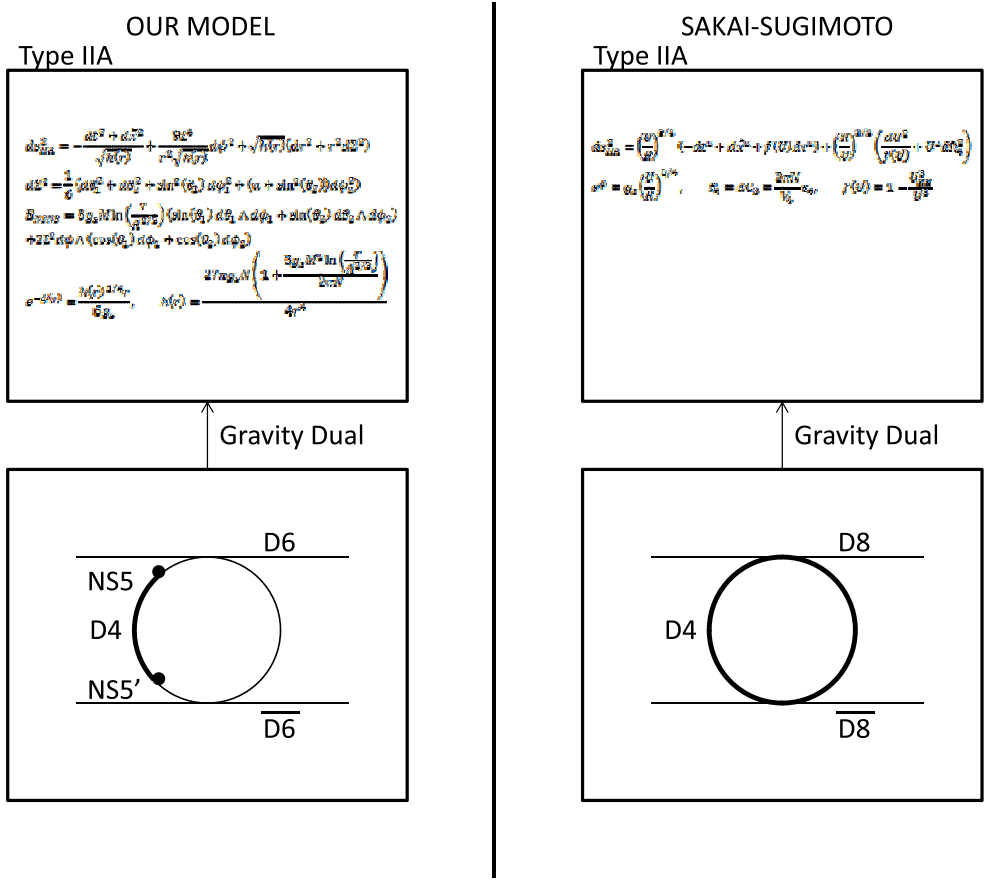}
\caption{Comparison with the type IIA Sakai-Sugimoto model.}
\label{compSS}
\end{figure}

\section{Comparision with the Sakai-Sugimoto model \label{SS}}

Let us take a short detour now to compare our model with the one constructed by Sakai and Sugimoto \cite{SS}. The Sakai-Sugimoto model is a beautiful brane construction in type IIA theory, which can be used to study certain IR dynamics of QCD, especially the scalar and vector mesonic spectra, via its gravity dual\footnote{The mapping of the spectra to real IR QCD is tricky, especially for the scalar mesonic sector because of certain ${\bf Z}_2$ parity. We will discuss this further later. See also \cite{SSrecent}.}. 
The model consists of a set of $N$ wrapped D4-branes on a circle giving the colors. The flavor D8 and \overbar{D8} branes are placed at the anti-podal points of the circle and respectively give rise to the $q \bar{q}$ mesonic pairs: the quark $q$ {arises from} {a} string {attaching {a} D8-brane to {the} set of D4-branes and the anti-quark $\bar{q}$ {arises similarly from} {a} string {attaching} {a} \overbar{D8}-brane to {the} set of D4-branes. The fluctuation spectrum of these strings are related to the mesonic spectrum.

In the gravity dual, the wrapped D4-branes are replaced by a geometry, i.e an asymptotically AdS space, but the eight-branes remain and so does the circular direction. The fluctuation spectrum of the quark and the anti-quark bound states is related to the various fluctuation modes of the eight-brane\footnote{The D8 and the \overbar{D8} branes combine to form one single U-shaped brane.} in the asymptotically AdS background. For example, fluctuations {\it parallel} to the eight-brane contribute to the vector mesonic spectrum, and the fluctuations {\it orthogonal} to the eight-brane are related to the scalar mesonic spectrum. 

The Sakai-Sugimoto model does not have a UV completion, so the comparison to our model can only be done in the 
IR\footnote{The Sakai-Sugimoto model treats the D8-branes as probes in an asymptotically AdS backgound. In this
language there is no issue about UV completion. But our model doesn't treat
such heavy objects as probes. So if we do not consider the D8-branes as probes, then
the gravitational effects of the D8 branes will go as $1 + \alpha|r|$, i.e linear
in $r$ with $\alpha$ some model dependent constant. This will blow up at large $r$. 
Note that this will persist even for $g_sN_f << 1$ as 
$g_s = {\cal O}(\epsilon)$ and $r = {\cal O}(1/\epsilon)$ for large $r$. In our IIB model, we have D7-branes
where the gravitational effects go as $1 + \beta{\rm log}~ r$. This too blows up, but we
inserted anti D7-branes, to kill the ${\rm log}~ r$ behavior and keep the $1/r$
behavior (i.e remove the {\it monopole} part and keep only the {\it dipole} part).
We also killed the tachyons by switching on appropriate world-volume fluxes.
This way our IIB model has no Landau poles or UV divergences. In our
understanding, no paper so far has touched this topic in details as most treatments on the subject keep such heavy objects as probes. This way all these models are {\it not} UV complete in our language.}. 
In the IR, our model differs considerably with the Sakai-Sugimoto construction. This is depicted in {\bf figure \ref{compSS}}. The differences can be elaborated in the following way.

\vskip.1in

\noindent $\bullet$ The $M$ D4-branes in our model are placed between two NS5-branes that are orthogonal to each other. Therefore, the D4-branes do not wrap the full $\psi$ circle. In the presence of type IIB fluxes, the T-dual NS5-branes' gravitational attraction is balanced by the charge repulsion between them. The strong coupling limit is given by the two orthogonal NS5-branes on top of each other. The theory on the brane is then exactly $3+1$ dimensional $SU(M)$ QCD\footnote{This construction should also be reminiscent of the geometric transition \cite{vafaGT} way of approaching the gravity dual whose brane construction was given in \cite{gt}.}. 

\vskip.1in

\noindent $\bullet$ The supersymmetry in our model is broken by both the embedding and the fluxes. However, the IIB and IIA models {are} stable as we discussed earlier. In particular, there are no tachyonic instabilities in our model and all gravitational  attractions are balanced by RR or NS charge repulsions.

\vskip.1in

\begin{figure}[H]
\centering
\includegraphics[width=\textwidth]{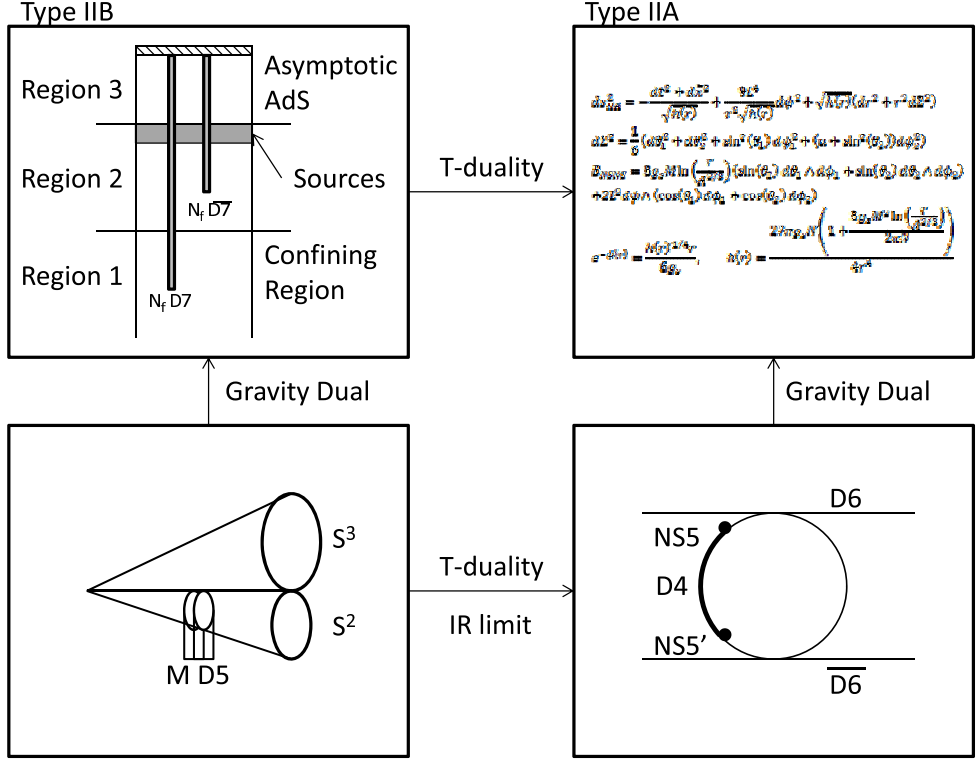}
\caption{The type IIA dual configuration.}
\label{dualdiagram}
\end{figure}

\noindent $\bullet$ The $N_f$ flavor branes in IIA are given by the D6-branes (and the \overbar{D6}-branes). These are T-dual to the $N_f$ D7 and the \overbar{D7} branes in the type IIB side. The $SU(N_f) \times SU(N_f)$ global symmetry is broken to $SU(N_f)$ in two possible ways in our model: one via the usual D7 and the \overbar{D7} branes combining in Region 1 to form a U-shaped D7-brane and the other one via the embedding depicted in {\bf figure \ref{dualdiagram}}. For our purpose, we will study a single, i.e $N_f = 1$, flavor D6-brane along with its parallel and orthogonal fluctuations. 

\vskip.1in

\noindent $\bullet$ Our type IIA gravity dual has a very different geometry from what considered by Sakai-Sugimoto so the subsequent analysis will be different from theirs. Since the AdS geometry shows up only in the large $r$ region in the type IIB picture, which in the language of the gauge theory is the UV region, we can probably compare the UV physics.

\vskip.1in

\noindent $\bullet$ At high temperatures, the issues raised by Mandal-Morita \cite{manmor} will clearly be absent in our set-up. This is because, since the distance between the two NS5-branes in the type IIA picture can be made arbitrarily small, it will require  very high energies to excite vibrations parallel to the $\psi$ direction of the D4-branes. Thus, the D4-branes inside an interval of the $\psi$ circle will effectively reproduce a $3+1$ dimensional QCD to arbitarily high energies. This in turn implies that {\it black} D4-branes will be able to capture the high-temperature physics of the model and the confinement to deconfinement phase transition will be given by the transition of solitonic D4-branes to black D4-branes. The Mandal-Morita's Gregory-Laflamme transitions will not play any role here.

\vskip.1in

\noindent Our complete analysis therefore is depicted in {\bf figure \ref{dualdiagram}}. On the bottom left of the figure, we have the IR brane configurations where we have wrapped D5-branes on a vanishing two-cycle of the conifold. The full resolved conifold picture, capturing the UV to IR physics, is already depicted in {\bf figure \ref{condet}}. The gravity dual in type IIB is given in the upper left box where we have clearly marked the three regions. Of course, the upper left box is the gravity dual of {\bf figure \ref{condet}}, but for simplicity we depict the IR brane configuration only. This is because a T-duality along the $\psi$ direction will give us the type IIA picture on the lower right-hand box in {\bf figure \ref{dualdiagram}}. This is the IR story and the comparison to Sakai-Sugimoto appears in {\bf figure \ref{compSS}}. The IR gravity dual in type IIA that will be studied throughout this paper is presented in the top right-hand box in {\bf figure \ref{dualdiagram}}. This gravity dual, which is of course the T-dual of the type IIB side for small $r$ regions, will be used to compute the vector and scalar mesonic spectra. In the following, we will start with the vector meson action.

\section{Vector Meson Action \label{secvec}}

The vector mesons arise from a gauge flux ($A_M$) that we insert along the Minkowski and $Z$ directions.
\begin{align}
A_M &= 
\begin{cases}
A_\mu(x^\mu,Z) \quad &\text{when $M = \mu \in \{t,x,y,z\}$} \\
A_Z(x^\mu,Z) \quad &\text{when $M = Z$} \\
0 \quad &\text{when $M \in \{\theta_2,\phi_2 \}$}
\end{cases} 
\end{align}
The gauge fluctuation is abelian and therefore we 
have the usual definition for the field strength $F_{MN}$:
\begin{equation}
F_{M N} = \partial_M A_N - \partial_N A_M
\end{equation}
This adds an extra abelian gauge field in the DBI action. Keeping terms quadratic in $F_{M N}$ in the DBI action, we have:
\begin{align}
S_{D6} &= -T \int d^4x \, dZ \, d\theta_2 d\phi_2 \, e^{-\phi(r(0,Z))} \sqrt{-\text{det}(g_6 + B_6 + 2 \pi \alpha' F)} \notag\\
&= - (2 \pi \alpha')^2 T \int d^4x \, dZ \,\left( v_1(Z) \, \eta^{\mu \nu} F_{\mu Z} F_{\nu Z} +   v_2(Z) \, \eta^{\mu \nu} \eta^{\rho \sigma} F_{\mu \rho} F_{\nu \sigma} + \ldots\right) \label{EQ:VMDBIaction}
\end{align}
where $d^4x \equiv dt\, dx\, dy\, dz$ and, $v_1(Z)$ and $v_2(Z)$ are coefficients whose values depend on our choices of the pull-backs $g_6$ and $B_6$ of the metric and $B$-field respectively. They are given by:
\begin{align}\label{vdeff}
v_1(Z) &\equiv \frac{\pi^2 e^{-\phi (r(0,Z))} \left(\cA^{8/3} \left(\pi ^2-24 (1+\squ)\right) e^{4 \left| Z\right| } h(r(0,Z))+81 \left(\pi ^2-24\right)(g_s M Z)^2\right)}{144 \, h(r(0,Z))^{3/4} \sqrt{\cA^4 (1+\squ) e^{6 \left| Z\right| } h(r(0,Z))+81 \cA^{4/3} e^{2 \left|Z\right| } (g_s M Z)^2}} \nonumber\\
v_2(Z) &\equiv \frac{1}{2} \cA^{4/3} e^{2 \left| Z\right| } h(r(0,Z)) \, v_1(Z) 
\end{align}
Before proceeding further, first note that in the 
presence of the squashing factor $\squ$, the $\theta_2$ integral becomes elliptical. We simplify this issue by first expanding the integrand around $\theta_2 = \pi/2$ up to second order in $\theta_2$ before performing the integration. We find that going to higher order in $\theta_2$ slightly improves the end results for a given value of $\squ$ and $\delta$ (see section \ref{sec6}). Therefore, we omit to consider higher-order terms to simplify the 
calculations. 
\\

The type IIA dilaton \eqref{dilla} has the following behavior:
\begin{equation}
e^\phi = 6g_s \left({4\over 27\pi g_s N}\right)^{1/4}\left(1-{3g_sM^2\over 8\pi N}\rho\right)
\end{equation}
{If we use the prescription mentioned in footnote 23 of \cite{FEP}}:
\begin{equation}
g_s \to \epsilon^{5/2}, ~~~~ {g_s M^2\over N} \to \epsilon^{9/2}, ~~~~ g_sN \to \epsilon^{-11/2}, ~~~~ 
\rho \to \epsilon^{-1}
\end{equation}
{$e^\phi \approx \epsilon^4$ and it is hence well behaved for all of Region 1.}
Our next set of steps are somewhat identical to the ones considered in Sakai-Sugimoto \cite{SS}, although the specific details will differ. Our {main interest lies in the} study {of} the gauge field fluctuations along the space-time directions. {To separate the space-time and $Z$ direction fluctuations (the ($\theta_2, \phi_2$) fluctuations are not considered here), we decompose the fluctuations of the $Z$-direction with specific {\it eigenmodes}.} These eigenmodes are exactly the stringy $q \bar{q}$ vibrational modes giving rise to the vector mesonic spectra. 

{To achieve this eigenmode expansion, we use} two sets of eigenfunctions $\{\alpha_n(Z), n \ge 1\}$ and $\{\beta_n(Z), n \ge 0\}$ whose orthogonality conditions and eigenvalue equation are defined later:
\begin{align}
A_\mu(x^\mu,Z) &= \sum_{n=1}^{\infty} B_{\mu}^{(n)}(x^\mu)\alpha_n(Z)\\
A_Z(x^\mu,Z) &= \sum_{n=0}^{\infty} \varphi^{(n)}(x^\mu)\beta_n(Z)\label{EQ:Amu} 
\end{align}
The field strengths of these gauge fluctuations can also be expressed in terms of the eigenmodes as:
\begin{align}
F_{\mu \nu} &= \sum_{n=1}^{\infty} \left(\partial_\mu B_{\nu}^{(n)}(x^\mu) - \partial_\nu B_{\mu}^{(n)}(x^\mu) \right)\alpha_n(Z) \equiv \sum_{n=1}^{\infty} F_{\mu \nu}^{(n)} \alpha_n(Z) \notag \\
F_{\mu Z} &= \partial_\mu \varphi^{(0)}(x^\mu) \beta_0(Z) +  \sum_{n=1}^{\infty} \left(\partial_\mu \varphi^{(n)}(x^\mu) \beta_n(Z) - B_{\mu}^{(n)}(x^\mu)\dot{\alpha}_n(Z) \right) 
\end{align}
Focusing on terms quadratic in $\alpha_n$ and $\beta_n$ (and forgetting about $\beta_0$ for now), we obtain action terms reminiscent of the vector meson terms of QCD:
\begin{align}
S_{\alpha_n^2,\beta_n^2} &= - (2 \pi \alpha')^2 T \int d^4x \, dZ \sum_{m,n}
 \Bigg[  v_2(Z) \, F_{\mu \nu}^{(n)} F^{\mu \nu (m)} \alpha_n \alpha_m + v_1(Z) B_{\mu}^{(m)} B^{\mu (n)}\dot{\alpha}_m\dot{\alpha}_n \\
&\qquad + v_1(Z) \left( \partial_\mu \varphi^{(n)} \partial^\mu \varphi^{(m)} \beta_n \beta_m - 2B_\mu^{(m)} \partial^\mu \varphi^{(n)} \dot{\alpha}_m \beta_n \right)\Bigg] 
\end{align}
where $v_i(Z)$ are given in \eqref{vdeff}. In order to recover the typical coefficient of the mass term, the $\alpha_n$ eigenfunctions must satisfy the identity:
\begin{align}
(2 \pi \alpha')^2 T \int \, dZ \, v_1(Z)  \dot{\alpha}_m \dot{\alpha}_n &= \frac{1}{2} \, m_n^2 \delta_{m n} 
\label{EQ:dalphacond}
\end{align}
Such identity is obtained by imposing the following orthogonality condition and eigenvalue equation:
\begin{align}
& (2 \pi \alpha')^2 T \int \, dZ \, v_2(Z) \, \alpha_m \alpha_n = \frac{1}{4}\delta_{mn} \label{EQ:alphanormcond}\\
&\partial_Z \left( v_1(Z) \, \partial_Z \alpha_n \right) + 2 \, v_2(Z) \, m_n^2 \alpha_n = 0 \label{EQ:eeq}
\end{align}
where $m_n^2 \equiv \lambda_n \mathcal{M}^2$ is the effective squared-mass of each vector meson and $\lambda_n$ is the eigenvalue of the corresponding mode. The mass scale $\mathcal{M}^2$ is given by $\frac{{\cal A}^{4/3}}{4 \pi g_s N}$.

Regarding the treatment of the $\beta_n$ eigenfunctions, we borrow some of the arguments used by Sakai and Sugimoto \cite{SS}. In order to normalize the kinetic term $\partial_\mu \varphi^{(n)} \partial^\mu \varphi^{(m)}$ to its canonical form, we must impose the following normalization condition for $\beta_n$:
\begin{align}\label{betacondi}
(2 \pi \alpha')^2 T \int\limits_{-\infty}^{\infty} \, dZ \, v_1(Z)  \beta_m \beta_n &= \frac{1}{2} \delta_{m n}
\end{align}
It is easily seen that choosing $\beta_n \equiv \frac{\dot{\alpha}_n}{m_n}$  for $n \ge 1$ will give us the necessary condition since we have (\ref{EQ:dalphacond}). Also, assuming that we are in a gauge where $\Lim{Z \rightarrow \pm \infty} A_{\mu}(Z) = 0$, we set $\beta_0 \equiv \frac{K}{v_1(Z)}$ for some normalization constant $K$. The fact that $A_{\mu}(Z)$ asymptotically vanishes guarantees orthogonality between $\beta_0$ and $\dot{\alpha}_n, \forall n \ge 1$.
\begin{align}
(2 \pi \alpha')^2 T  \int \, dZ \, C(Z)  \beta_0 \dot{\alpha}_n &= (2 \pi \alpha')^2 T K \int_{-\infty}^{\infty} \, dZ \, \partial_Z \alpha_n = 0
\end{align}
Going back to our expression for $F_{\mu Z}$, we have:
\begin{align}
F_{\mu Z} &= \partial_\mu \varphi^{(0)} \beta_0(Z) +  \sum_{n=1}^{\infty} \left(m_n^{-1}\partial_\mu \varphi^{(n)}  - B_{\mu}^{(n)} \right)\dot{\alpha}_n(Z)
\end{align}
Absorbing $m_n^{-1} \partial_\mu \varphi^{(n)}$ into $B_{\mu}^{(n)}$ with the gauge transformation:
\begin{align}
B_{\mu}^{(n)} \rightarrow B_{\mu}^{(n)} + m_n^{-1} \partial_\mu \varphi^{(n)}
\end{align}
we obtain typical meson terms in the QCD action where $\varphi^{(0)}(x^\mu)$ is the Nambu-Goldstone boson of the broken chiral symmetry:
\begin{align}
S_{\text{QCD, Vector}} = - \int d^4x \left[\frac{1}{2}\partial_\mu \varphi^{(0)} \partial^\mu \varphi^{(0)} + \sum_{n = 1}^{\infty}
 \left(\frac{1}{4} F_{\mu \nu}^{(n)} F^{\mu \nu (n)}
 + \frac{1}{2} m_n^2 B_{\mu}^{(n)} B^{\mu (n)} \right) \right]
\end{align}

\section{Vector Meson Spectrum \label{S:VM}}

We now solve the eigenvalue equation (\ref{EQ:eeq}) by using simple perturbation techniques with $\delta \equiv \frac{g_s M^2}{N}$ as the controlling parameter. We introduce some notation to write the problem in terms of a differential operator $\mathbf{H}_{\text{v}}$ acting on its eigenfunctions $\alpha_n$. For example, we can recast \eqref{EQ:eeq} in terms of an eigenvalue equation in the following way: 
\begin{equation}
(\ref{EQ:eeq}) \rightarrow  \mathbf{H}_\text{v} | \alpha_n \rangle =  \lambda_n |\alpha_n\rangle  \label{EQ:qmeeq}
\end{equation}
where the Hamiltonian $\mathbf{H}_{\rm v}$ is defined by looking at \eqref{EQ:eeq}:
\begin{equation}
\mathbf{H}_{\text{v}} \equiv  - \frac{v_1(Z)}{2 \, \mathcal{M}^2 v_2(Z) } \left( \partial_Z^2 + \frac{v_1'(Z)}{v_1(Z)}  \, \partial_Z   \right)
\end{equation}
Once we have an eigenvalue equation of the form \eqref{EQ:qmeeq}, the states $\vert\alpha_n\rangle$ have to be orthogonal. We use $f(Z)$:
\begin{equation}
f(Z) \equiv 4 \,(2 \pi \alpha')^2 T \, v_2(Z)
\end{equation}
to define the orthogonality condition in the following way:
\begin{equation}\label{ctota}
\langle \alpha_m | \alpha_n \rangle \equiv \int\limits_{-\infty}^{\infty} dZ f(Z) \alpha_m \alpha_n = \delta_{mn} 
\end{equation}
The ket states $\vert\alpha_n\rangle$ of course have the usual representations in terms of the Schr\"odinger states:
\begin{equation}
\langle Z \vert \alpha_n\rangle = \alpha_n(Z)
\end{equation}
and we expand both the states $\alpha_n$ as well as the corresponding eigenvalues $\lambda_n$ order by order 
in the expansion parameter $\delta$:
\begin{align}
\alpha_n &= \alpha_n^{(0)} + \delta \, \alpha_n^{(1)} + \delta^2 \alpha_n^{(2)} + \ldots \nonumber\\
\lambda_n &= \lambda_n^{(0)} + \delta \, \lambda_n^{(1)} + \delta^2 \lambda_n^{(2)} + \ldots
\end{align}
The above expansions are consistently realized once the Hamiltonian $\mathbf{H}_{\rm v}$ is also expanded in terms of $\delta$:
\begin{equation}
\mathbf{H}_{\text{v}} = \mathbf{H}_{\text{v}}^{(0)} + \delta\, \mathbf{H}_{\text{v}}^{(1)} + \delta^2 \mathbf{H}_{\text{v}}^{(2)} + \ldots
\end{equation}
Performing the same expansion on the orthogonality function $f(Z)$, we can express \eqref{ctota} as:
\begin{align}
\langle \cdot | \cdot \rangle &= \int\limits_{-\infty}^{\infty} dZ f^{(0)}(Z) + \delta  \int\limits_{-\infty}^{\infty} dZ f^{(1)}(Z) + \delta^2  \int\limits_{-\infty}^{\infty} dZ f^{(2)}(Z) + \ldots \\
&\equiv \langle \cdot| \cdot \rangle^{(0)} + \delta \, \langle \cdot| \cdot \rangle^{(1)} + \delta^2 \langle \cdot| \cdot \rangle^{(2)} + \ldots
\end{align}

\subsection{Zeroth-Order Eigenvalues and Eigenfunctions \label{seczero}}

It is now easy to work out the eigenvalue equation order by order in $\delta$. 
Solving eq. (\ref{EQ:qmeeq}) in $\delta$, we first have\footnote{We assume that the identification $\text{sign}(Z) = \frac{Z}{|Z|}$ is valid for all values of Z. This equality is in general incorrect at $Z = 0$.  But since the point $Z = 0$ has zero measure, the values of the Z integrals and the eigenvalues should not change. }:
\begin{align}
\mathbf{H}_{\text{v}}^{(0)} | \alpha_n^{(0)} \rangle &= \lambda_n^{(0)} | \alpha_n^{(0)} \rangle \quad \text{with} \quad \langle \alpha_m^{(0)} | \alpha_n^{(0)} \rangle^{(0)} = \delta_{mn} \label{EQ:normcond} 
\end{align}
where the zeroth-order Hamiltonian $\mathbf{H}_{\rm v}^{(0)}$ and the function $f^{(0)}$ are given by the following expressions:
\begin{align}
\mathbf{H}_{\text{v}}^{(0)} &= - e^{2 |Z|} \left( \partial_Z^2 + 2 \, \text{sgn}(Z) \, \partial_Z   \right) \\
f^{(0)} &= T \alpha'^2 \frac{\left(1 + \squ - \frac{\pi^2}{24}\right)}{\sqrt{1+\squ}}\left(\frac{3 N^3 \pi^{19}}{4 g_s}\right)^{1/4}
\end{align}
The differential operator is invariant under $Z \rightarrow - Z$, which allows us to characterize the eigenfunctions as either odd or even. The zeroth-order eigenvalue equation is:
\begin{align}
\ddot{\alpha}_n^{(0)}(Z)& + 2 \, \text{sign}(Z) \dot{\alpha}_n^{(0)}(Z)  + e^{- 2|Z|}  \lambda_n \alpha_n^{(0)}(Z) = 0 \label{EQ:de}
\end{align}

To avoid the discontinuity at $Z = 0$, we solve the eigenvalue equation in two regimes, namely $Z > 0$ and $Z < 0$, with which we construct piecewise solutions of the full equation. The eigenfunctions are given in terms of Bessel's functions of the first and second kind.
\begin{align}
\alpha_n^{(0)}(Z) = C_n \, e^{-|Z|}  J_1\left(\sqrt{\lambda_n} e^{-|Z|}\right) + D_n \, e^{-|Z|}  Y_1\left(\sqrt{\lambda_n} e^{-|Z|}\right)
\end{align}
In order to satisfy the zeroth-order orthonormality condition, we must set $D_n = 0$ since $e^{-|Z|}  Y_1\left(\sqrt{\lambda_n} e^{-|Z|}\right)$ doesn't vanish at $Z \rightarrow \pm \infty$. $C_n$ are determined by using the zeroth-order normalization with the following manipulations:
\begin{align}
&\int_{-\infty}^{0^-} f^{(0)} (\alpha_n^{(0)}(Z))^2 dZ + \int_{0^+}^{\infty} f^{(0)} (\alpha_n^{(0)}(Z))^2 dZ = 1 \nonumber\\
&f^{(0)} \left(\int_{-\infty}^{0^-} (\alpha_n^{(0)}(Z))^2 dZ + \int_{0^+}^{\infty} (\alpha_n^{(0)}(Z))^2 dZ \right) = 1 \nonumber\\
&f^{(0)} C_n^2 \left( {J_1}^2\left(\sqrt{\lambda_n} \right) - J_0\left(\sqrt{\lambda_n}\right) J_2\left(\sqrt{\lambda_n}\right)\right) = 1 \nonumber
\end{align}
from which we can deduce an expression for $C_n$ in terms of the Bessel's functions: 
\begin{equation}
C_n = \frac{1}{\sqrt{f^{(0)}}}\frac{1}{\sqrt{\left( {J_1}^2\left(\sqrt{\lambda_n} \right) - J_0\left(\sqrt{\lambda_n}\right) J_2\left(\sqrt{\lambda_n}\right)\right)}} 
\end{equation}
The eigenvalues are obtained by solving the following equations, which we expect for odd and even functions. These conditions also guarantee perfect orthonormality of the eigenfunctions (see eq. \ref{EQ:OrthCond}).
\begin{align}
\alpha_n^{(0)}(0,\lambda_n) = J_1\left(\sqrt{\lambda_n} \right) &= 0\quad (\text{Odd functions})\label{EQ:oddcond}\\
\partial_Z \alpha_n^{(0)}(0,\lambda_n) = J_0\left(\sqrt{\lambda_n} \right) &=0 \quad (\text{Even functions})\label{EQ:evencond}
\end{align}
The eigenvalues can then be read graphically as depicted in {\bf figure \ref{zeroed}} below.
\begin{figure}[H]
\centering
\begin{subfigure}{0.48\textwidth}
\includegraphics[width=\textwidth]{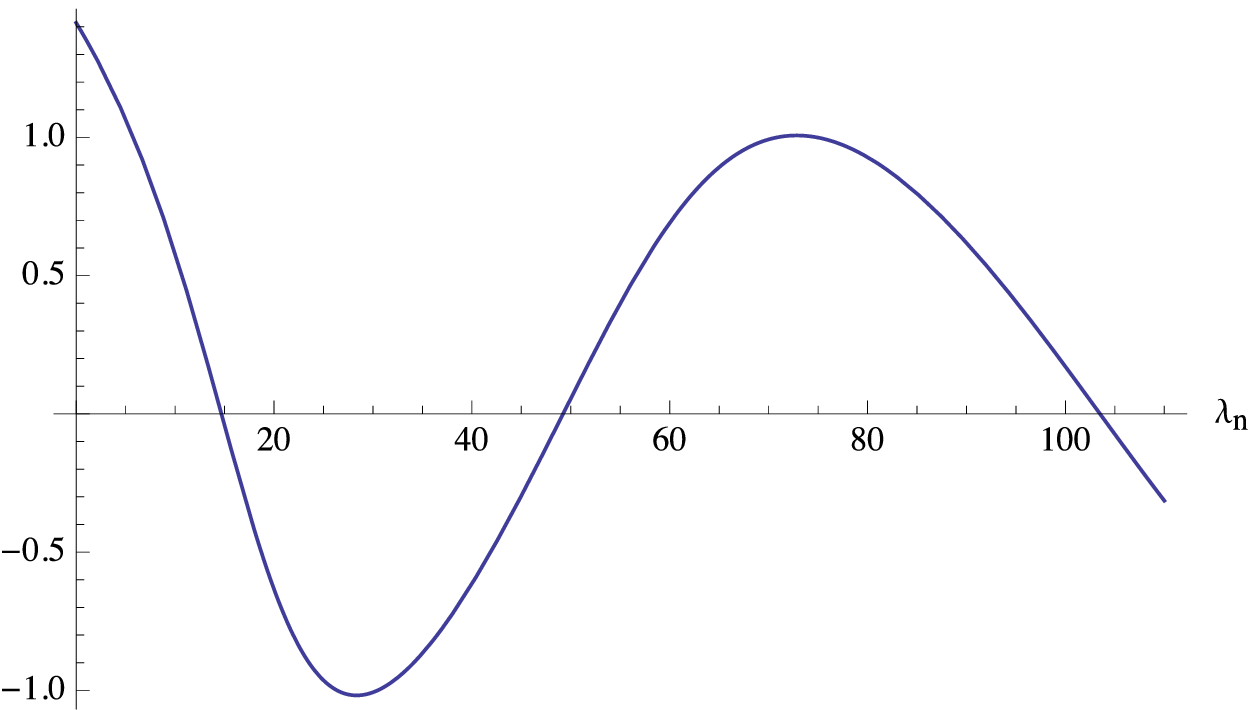}
\subcaption{$\alpha_n(0, \lambda_n)$}
\end{subfigure}
\begin{subfigure}{0.48\textwidth}
\includegraphics[width=\textwidth]{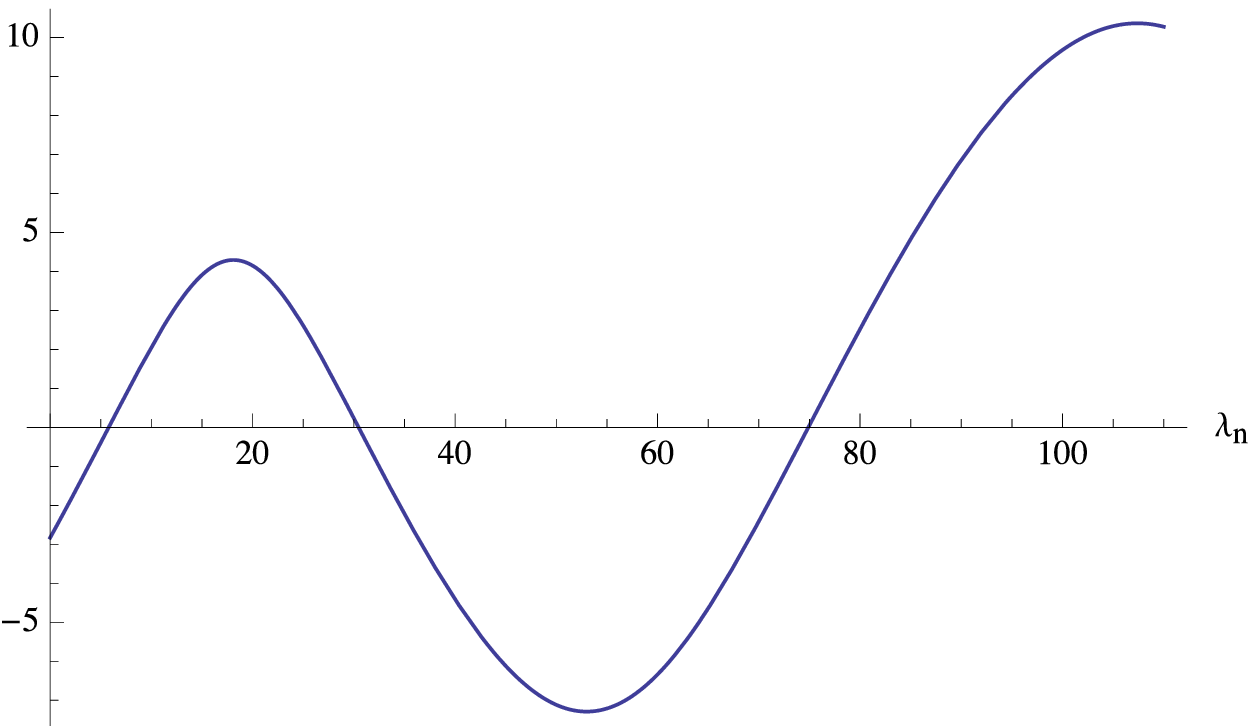}
\subcaption{$\partial_Z \alpha_n(0,\lambda_n)$}
\end{subfigure}
\caption{Zeroes of $\alpha_n(0, \lambda_n)$ and $\partial_Z \alpha_n(0,\lambda_n)$.}
\label{zeroed}
\end{figure}
For odd functions, we also add an extra $\text{sign}(Z)$ to make them truly odd. Using the same indexing as Sakai \& Sugimoto, we have the following summary:
\begin{align}
&\alpha_{2n + 1}^{(0)}(Z) = C_{2n+1} \, e^{-|Z|}  J_1\left(\sqrt{\lambda_{2n+1}} e^{-|Z|}\right) \nonumber\\
&\alpha_{2n}^{(0)}(Z) = C_{2n} \, \text{sign}(Z) \, e^{-|Z|}  J_1\left(\sqrt{\lambda_{2n}} e^{-|Z|}\right)
\end{align}

As we can see in the plots of {\bf figure \ref{normalizability}}, the even-indexed functions are truly odd after adding the extra $\text{sign}(Z)$. All the eigenfunctions are continuous, but fail to be differentiable at $Z = 0$ because of the functions $|Z|$ and $\text{sgn}(Z)$. 

Also, with the Bessel's functions, we obtain perfect orthogonality of the eigenfunctions. First, by introducing this extra $\text{sgn}(Z)$ in odd functions, one automatically obtains orthogonality between odd and even functions since the integral of the orthogonality condition has a symmetric range. Orthogonality between two odd or two even functions is also achieved by looking at the result of the integral in such cases and remembering eq. (\ref{EQ:oddcond}) and (\ref{EQ:evencond}).
\begin{align}
\langle \alpha_m^{(0)} | \alpha_n^{(0)} \rangle^{(0)} &= \int\limits_{-\infty}^{\infty} dZ f^{(0)}(Z) \alpha_m^{(0)}(Z) \alpha_n^{(0)}(Z) \quad \text{$m \ne n$, $(m,n) \in 2 \mathbb{Z}$ or $2 \mathbb{Z} + 1$ }\nonumber\\
&= \frac{f^{(0)} C_n C_m}{\lambda_m - \lambda_n} \left( \sqrt{\lambda_n} J_0(\sqrt{\lambda_n})J_1(\sqrt{\lambda_m})- \sqrt{\lambda_m} J_0(\sqrt{\lambda_m})J_1(\sqrt{\lambda_n})\right) \label{EQ:OrthCond}
\end{align}
The fact that we have perfect orthogonality is no surprise since the differential equation (\ref{EQ:de}) can be cast into a Sturm-Liouville equation, which guarantees the existence of a complete set of orthonormal eigenfunctions.

\begin{figure}[H]
\begin{tabular}{|c|c|c|}
\hline
\begin{subfigure}[H]{0.3\textwidth}
\vspace{0.1 cm}
\includegraphics[width=\textwidth]{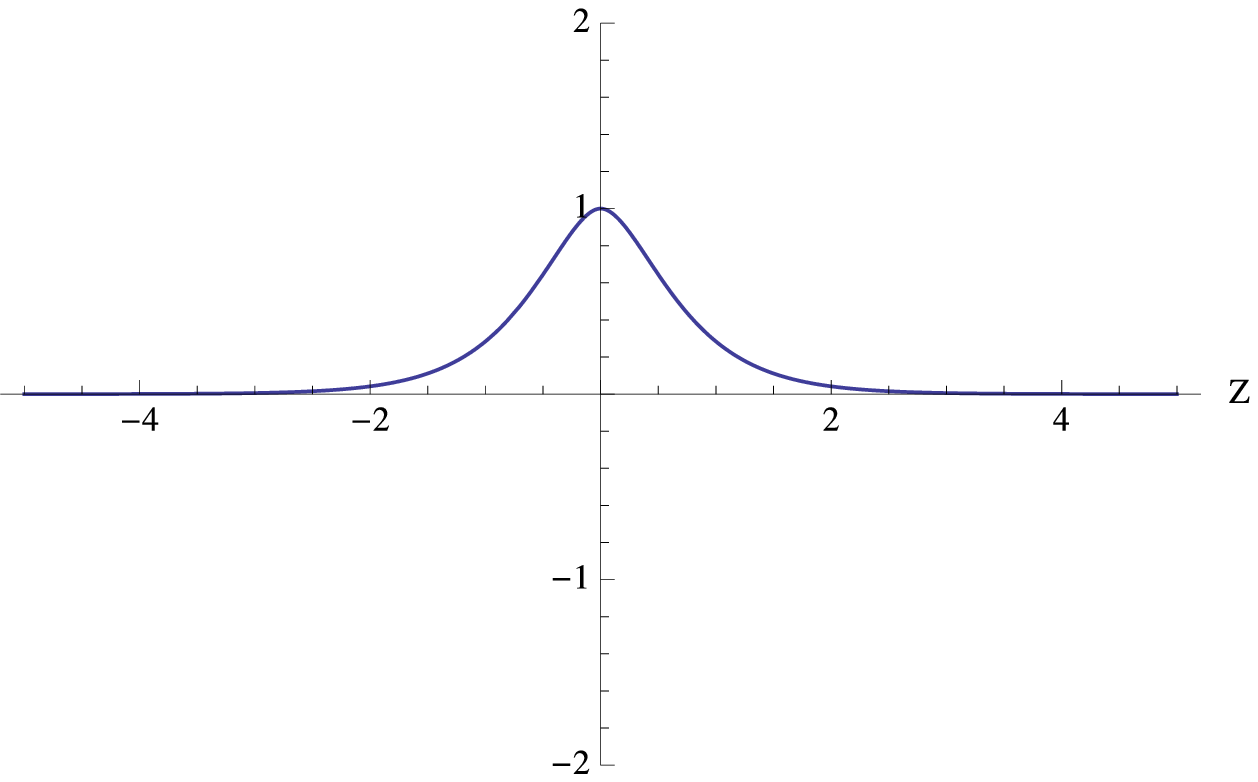}
\subcaption{$\alpha_1(Z)$}
\end{subfigure}
&
\begin{subfigure}[H]{0.3\textwidth}
\vspace{0.1 cm}
\includegraphics[width=\textwidth]{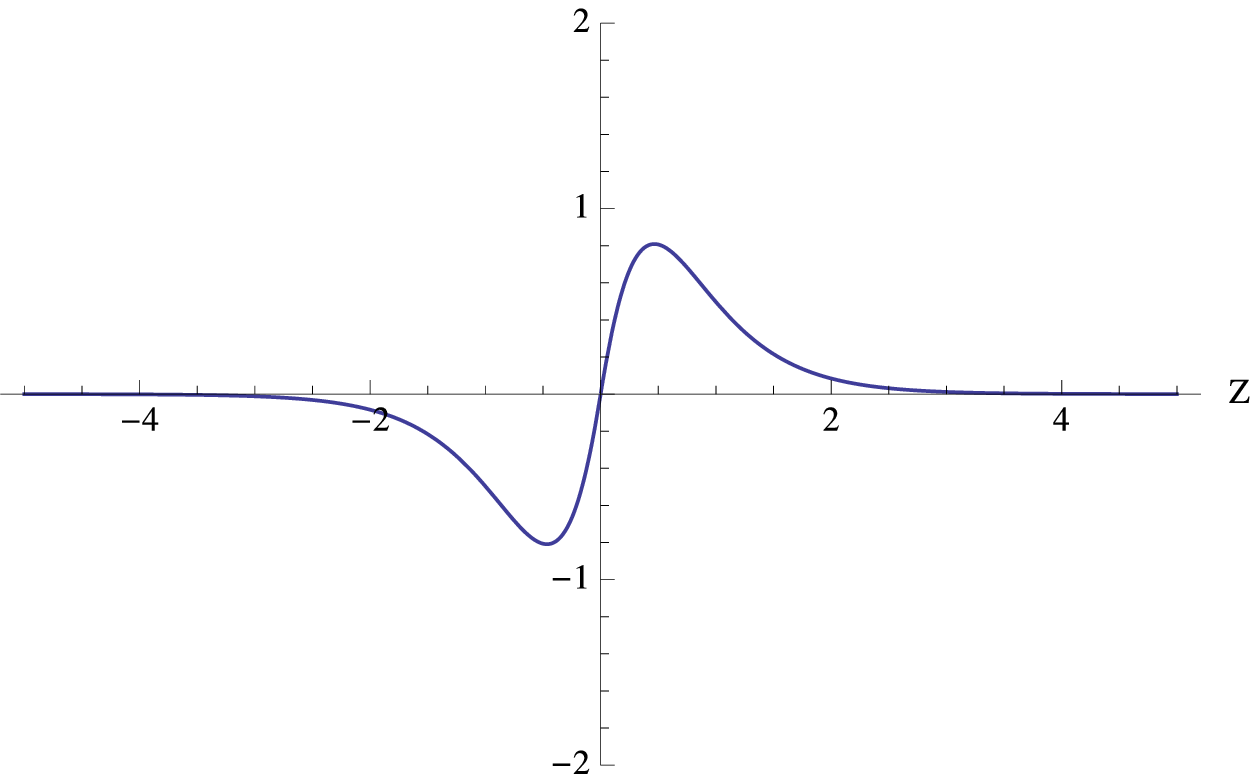}
\subcaption{$\alpha_2(Z)$}
\end{subfigure} 
&
\begin{subfigure}[H]{0.3\textwidth}
\vspace{0.1 cm}
\includegraphics[width=\textwidth]{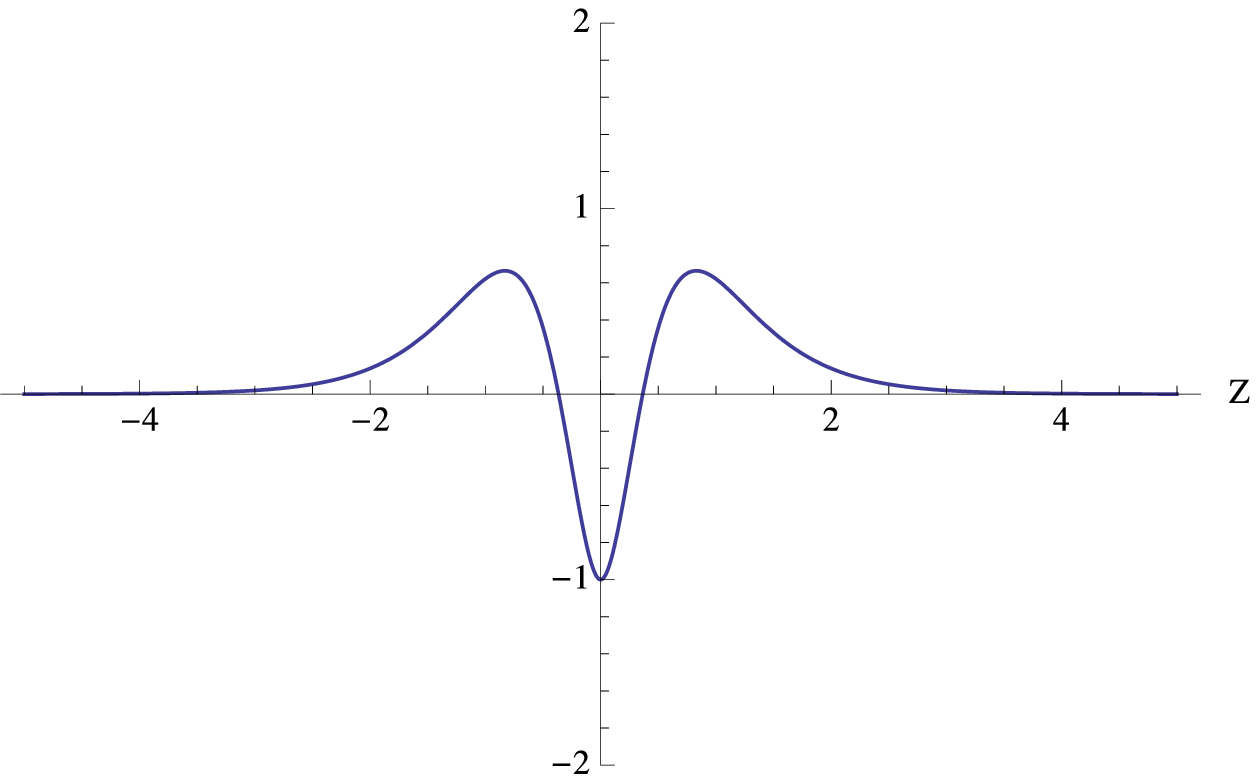}
\subcaption{$\alpha_3(Z)$}
\end{subfigure}\\
\hline
\begin{subfigure}[H]{0.3\textwidth}
\vspace{0.1 cm}
\includegraphics[width=\textwidth]{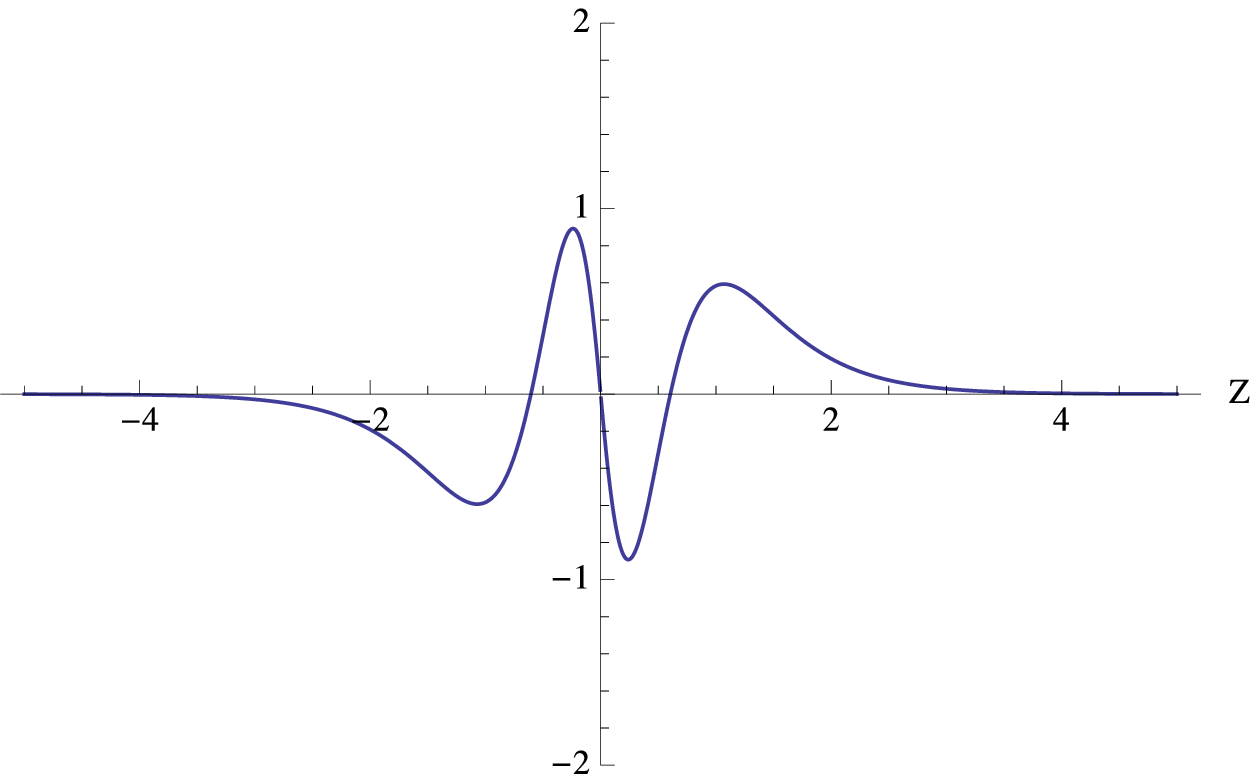}
\subcaption{$\alpha_4(Z)$}
\end{subfigure}
&
\begin{subfigure}[H]{0.3\textwidth}
\vspace{0.1 cm}
\includegraphics[width=\textwidth]{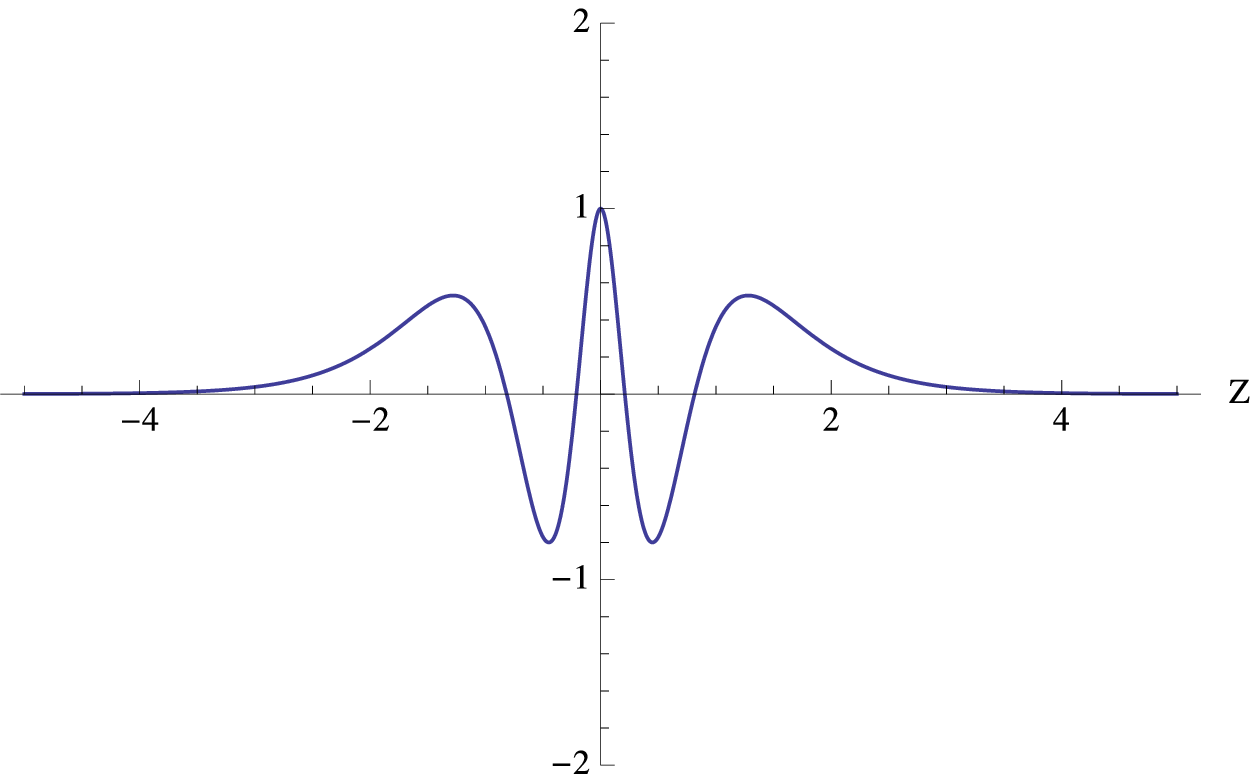}
\subcaption{$\alpha_5(Z)$}
\end{subfigure}
&
\begin{subfigure}[H]{0.3\textwidth}
\vspace{0.1 cm}
\includegraphics[width=\textwidth]{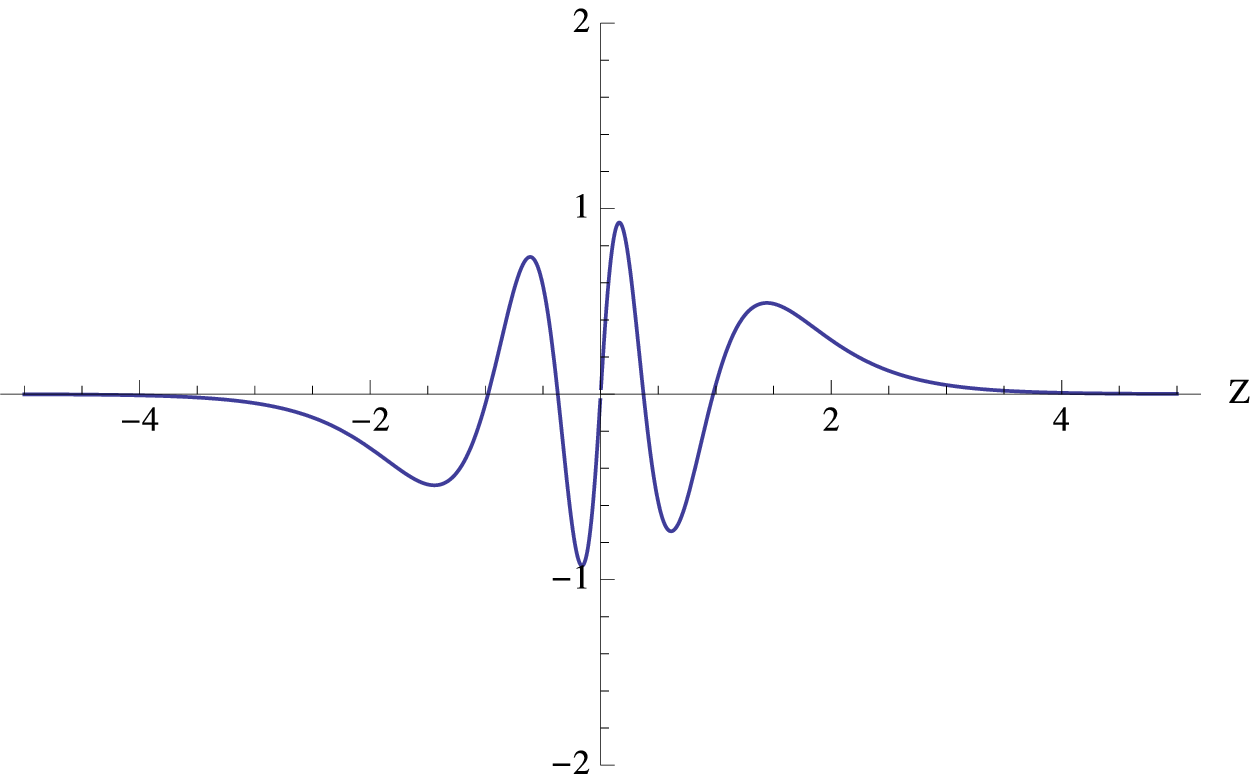}
\subcaption{$\alpha_6(Z)$}
\end{subfigure} \\
\hline
\end{tabular}
\caption{Zeroth-order eigenfunctions of the six lightest vector mesons.
For plotting purposes, we set the constant $f^{(0)} = 1$ since it is common to all of the eigenfunctions.}
\label{normalizability}
\end{figure}

\subsection{First-Order Eigenvalue \label{secfirst}}

We now assess the first-order correction to the eigenvalues of eq.(\ref{EQ:qmeeq}). The formula for such correction is well-known in the literature:
\begin{equation}
\lambda_n^{(1)} = \langle \alpha_n^{(0)}|\mathbf{H}_{\text{v}}^{(1)} | \alpha_n^{(0)} \rangle^{(0)} 
\end{equation}
where the first-order Hamiltonian $\mathbf{H}_{\rm v}^{(1)}$ is given by the following expression:
\begin{equation}
\mathbf{H}_{\text{v}}^{(1)} = \frac{3 \, e^{2 |Z|}}{2 \pi} \left[  |Z| \partial_Z^2 - 2\, Z \left(7 -\frac{4}{1+\squ} - \frac{192 \squ}{24 (1+\squ) - \pi^2}\right) \, \partial_Z \right] 
\end{equation}
After operating $\mathbf{H}_{\text{v}}^{(1)}$ and identifying $|Z|', |Z|''$ with $\text{sign}(Z),\, 2 \,\delta(Z)$ respectively, the integral evaluates to the following expression:
\begin{align}
\lambda_n^{(1)} &= \frac{3}{2 \pi \left({J_1}^2\left(\sqrt{\lambda_n }\right)-J_0\left(\sqrt{\lambda_n }\right) J_2\left(\sqrt{\lambda_n
   }\right)\right)} \bigg[ 1+ 2 \, \lambda_n  \,\left( \frac{24-\pi ^2+2 \squ \left(12-\pi ^2\right)}{(1+\squ) \left(24 +24 \squ -\pi ^2\right)} \right) {\bf F}_{3,4}  \notag\\
 &~~~~~~ - (1 + \lambda_n) {J_0}^2\left(\sqrt{\lambda_n }\right) - \lambda_n {J_1}^2\left(\sqrt{\lambda_n }\right) + \sqrt{\lambda_n } J_0\left(\sqrt{\lambda_n }\right) J_1\left(\sqrt{\lambda_n }\right)\bigg] \label{EQ:lambdan1}
\end{align}
where $\lambda_n$ stands for the zeroth-order eigenvalue $\lambda_n^{(0)}$ and ${\bf F}_{3,4}$
is a generalized hypergeometric function given as:
\begin{equation}
{\bf F}_{3,4} \equiv {}_3F_4\left(1, 1, {3\over 2}; 2, 2, 2, 2; -\lambda_n\right)
\end{equation}

\subsection{Meson Identification \label{secmeson}}

We would like to verify if this effective model of QCD shares even more similarities with the experimental model by comparing ratios of $m_n^2$ of well-known fields. In order to do so, we must first identify the kind of meson fields that are present in this effective theory by looking at their behavior under charge conjugation ($\mathcal{C}$) and parity ($\mathcal{P}$). 

First, we determine the eigenvalue ($\omega_{\mathcal{P},n} = \pm 1$) of the vector mesons $B_{\mu}^{(n)}$ upon action of the parity operator. In the 5-dimensional theory, the parity operator is a Lorentz transformation that flips the sign of spacelike coordinates.
\begin{align}
\mathcal{P} (x^0,x^1,x^2,x^3,Z) = (x^0,-x^1,-x^2,-x^3,-Z) \label{EQ:ParityOp}
\end{align}
Looking at the expansion of the four-dimensional gauge potential (\ref{EQ:Amu}), we conclude that $B_{\mu}^{(n)}$ must be odd (resp. even) under parity when $\alpha_n$ is even (resp. odd) in order for $A_\mu$ to behave as a 4-vector.

Second, the charge conjugation eigenvalue ($\omega_{\mathcal{C},n} = \pm 1$) is determined with a similar logic. From the point of view of the string theory, vector mesons are built from quarks, which correspond to modes of strings stretched between D4 and D6-branes (q) or D4 and \overbar{D6}-branes (\overbar{q}). Since the D6 and \overbar{D6} branes are antipodal on the $\psi$ cycle, by changing $\psi \rightarrow - \psi$ (or $Z \rightarrow - Z$) we interchange the position of the D6 and \overbar{D6} branes, which corresponds to changing the chirality of quarks in the effective QCD model. Hence, charge conjugation of the QCD model corresponds to a flip of the $Z$ coordinate. Looking again at eq. (\ref{EQ:Amu}), this means that $B_{\mu}^{(n)}$ must be odd (resp. even) under charge conjugation when $\alpha_n$ is even (resp. odd) since $A_\mu$ must acquire an overall sign under charge conjugation.

Knowing the eigenvalues of each vector mesons under $\mathcal{P}$ and $\mathcal{C}$, we can identify them using the Particle Data Group (PDG) database \cite{PDG} where we use their mass measurements $M_{\text{PDG}}$ for comparison. Also, we specify to fields that are vectors of the approximate isospin $SU(2)$ symmetry as it was clarified in \cite{Son:2003/04}. {One might argue that our meson identification has omitted the contribution of the {\it massive} open string modes to the vector masses, as addressed in \cite{SSrecent}.} We will discuss this in sec \ref{S:Sum}. 
The following table summarizes our knowledge of each vector mesons $B_{\mu}^{(n)}$.
\begin{table}[H]
\centering
\caption{Vector Mesons}
\begin{tabular}{| c | c | c | c | c | c | c | c |} 
\hline
 & $\lambda_n^{(0)}$ & $\alpha_n^{(0)}$ Parity &$\omega_{\mathcal{C},n}$ & $\omega_{\mathcal{P},n}$ & PDG Name & $M_{\text{PDG}}$(MeV) & Full Width $\Gamma_{PDG}$ (MeV) \\ \hline
$B_{\mu}^{(1)}$ & 5.78 & Even & + & + & $\rho(770)$ & 775.49 & 149.1\\ \hline
$B_{\mu}^{(2)}$ & 14.68 & Odd & - & - & $a_1(1260)$ & 1230 & 250\\ \hline
$B_{\mu}^{(3)}$ & 30.47 & Even & + & + & $\rho(1450)$ & 1465 & 400\\ \hline
$B_{\mu}^{(4)}$ & 49.22 & Odd & - & - & $a_1(1640)$ & 1647 & 254\\ \hline
$B_{\mu}^{(5)}$ & 74.89 & Even & + & + & $\rho(1700)$ & 1720 & 250\\ \hline
\end{tabular}
\end{table}

\section{UV Physics and contributions of the UV regions \label{UVphysics}}

Before moving further, let us discuss the contributions to the mesonic spectra of the UV (i.e large $r$) regions, {i.e. Regions 2 and 3}. 
The normalizability conditions for the $\alpha_n$ and $\beta_n$ modes are given in \eqref{EQ:alphanormcond} and 
\eqref{betacondi} respectively and are rewritten here as:
\begin{align}\label{norman}
&(2 \pi \alpha')^2 T \int \, dZ \, v_2(Z) \, \alpha_m \alpha_n = \frac{1}{4}\delta_{mn}\nonumber\\
&(2 \pi \alpha')^2 T \int \, dZ \, v_1(Z) \, \beta_m \beta_n = \frac{1}{2}\delta_{mn}
\end{align}
Incorporating the details of Regions 2 and 3, the above normalizability conditions should change to the following:
\begin{equation}
(2 \pi \alpha')^2 T \left(\int_{-\infty}^{-{\rm log}(r_{min}/r_0)} + \int_ {-{\rm log}(r_{min}/r_0)}^{+{\rm log}(r_{min}/r_0)} + \int_{+{\rm log}(r_{min}/r_0)}^{+\infty}\right)\, dZ \, v_2(Z) \, \alpha_m \alpha_n = \frac{1}{4}\delta_{mn}
\end{equation}
and similarly for the $\beta_n$ equation in \eqref{norman}. We can also assume that:
\begin{equation}
\int_{+{\rm log}(r_{min}/r_0)}^{+\infty}\, {\cal G}(Z; \alpha_n, \beta_m) dZ \equiv \left(\int_{{\rm log}(r_{min}/r_0)}^{{\rm log}(r_3/r_0)} + \int_{{\rm log}(r_3/r_0)}^{+\infty}\right)\, {\cal G}(Z; \alpha_n, \beta_m) dZ
\end{equation}
for a finer division of the UV regions of \eqref{norman} beyond $\rho \ge {\rm log}(r_{min}/r_0)$. Here ${\cal G}(Z; \alpha_n, \beta_m)$ stands for a generic function of $Z$ and the wave-functions ($\alpha_n, \beta_m$) on the gravity side.

The UV physics is relevant if the wave-functions $\alpha_n$ and $\beta_n$ slowly decay to zero in the regime $r > r_{min}$. In fact, {only} Region 2 {is} relevant if the decay of the wave-functions occurs in the regime $r_{min} < r < r_3$, {but} both Regions 2 and 3 {are} relevant if the decay starts at $r > r_3$. 

The above discussion gives us a way to see under what conditions the UV regimes become irrelevant: the wave-functions $\alpha_n$ and $\beta_n$ should decay fast enough to zero before we hit the boundary of Region 2, i.e before we reach $r = r_{min}$. Since both $v_1(Z)$ and $v_2(Z)$ defined in \eqref{vdeff} are non-normalizable functions, the above constraint on the wave-functions is essential for UV irrelevancy. In the limit of small $r_0$, large $r_{min}$ and wave-functions ($\alpha_n, \beta_n$) {localized} in Region 1, the replacement:
\begin{equation}
\int_ {-{\rm log}(r_{min}/r_0)}^{+{\rm log}(r_{min}/r_0)} {\cal G}(Z; \alpha_n, \beta_m) dZ ~~~ \to ~~~ \int_{-\infty}^{+\infty} {\cal G}(Z; \alpha_n, \beta_m) dZ
\end{equation}
{eliminates the UV physics by neglecting the contribution of Regions 2 and 3. This essentially means that the $\delta$ corrections of the wave-functions and the normalizability integral are almost exactly computed by considering the contributions of Region 1 {\it only}. This assumption is crucial because it significantly simplifies the analytic derivation of the spectrum's masses.} Alternatively, this means that the boundary ${\rm log}(r_{min}/r_0)$ is far enough to allow for higher order $\delta$ corrections to the wave-functions and yet keep the normalizability within the bounds of Region 1. What happens for the massive sector is another story that will be briefly discussed in sec 
\ref{S:Sum}. 

\section{Scalar Meson Action \label{secmesac}}
In our holographic QCD model, scalar mesons arise as fluctuations of the orthogonal directions of the probe D6-branes. This means that we could have fluctuations along the sphere parametrized by ($\theta_1, \phi_1$) or along $Y$. For our choice of embedding $\theta_1 = \phi_1 = 0$, it is difficult to consider brane fluctuations along the $\theta_1, \phi_1$ directions because of the intricate deformed-conifold geometry. We therefore restrict ourselves to $Y$ fluctuations only, which were also studied by Sakai \& Sugimoto, and determine the resulting DBI action focusing on terms quadratic in the field $Y(x^\mu,Z)$. The calculations for the scalar mesons' spectrum are similar to the vector mesons' spectrum with some additional subtleties. The full scalar mesonic action is written as:
\begin{equation}
S_{Y^2} = - T \int d^4x \, dZ \left[s_1(Z)\partial_\mu Y \partial^\mu Y +  s_2(Z) \dot{Y}^2 + s_3(Z) Y^2 \right] \label{EQ:SMDBIaction}
\end{equation}
where the coefficients $s_1(Z), s_2(Z)$ can be derived from the background data {and result in the following expressions}:
\begin{align}
s_1(Z) &= \frac{s_0 e^{2 \left| Z\right| } \left[2 N \pi  \left(72+72 \squ+\pi ^2\right)+3 g_s M^2 \left(24 \left(24-\pi ^2\right) Z^2+\left(72+72 \squ+\pi ^2\right) \left| Z\right| \right)\right]}{32 Z^2 \left(2 N \pi +3 g_s M^2 \left| Z\right| \right)^{3/4} \sqrt{2 (1+\squ) N \pi +3 g_s M^2 \left(8 Z^2+(1+\squ) \left| Z\right| \right)}}\nonumber\\ 
s_2(Z) &= \frac{8 \cA^{4/3} e^{2 \left| Z\right| }}{54 g_s N \pi +81 g_s^2 M^2 \left| Z\right| } \,\, s_1(Z), ~~~~~~~~ s_0 \equiv 3{\cal A}^{4/3} c^2 N \pi^3 \left({3g_s^3\over 2}\right)^{1/4}
\end{align}
with the mass-term function $s_3(Z)$ given by a much bigger expression in terms of the background data:
\begin{align}
s_3(Z) &= -\frac{\cA^{8/3} c^2 N \pi ^3 e^{4 \left| Z\right| }}{48 (54 g_s)^{1/4} \left| Z\right|^{5}  \left(2 N \pi +3 g_s M^2 \left| Z\right| \right)^{11/4} \left(2 (1+\squ) N \pi +3 g_s M^2 \left(8 Z^2+(1+\squ) \left| Z\right| \right)\right)^{3/2}} \notag\\
\times &\Bigg[ 64 (1+\squ) N^3 \pi ^3 \left(72+72 \squ+\pi ^2\right) \left(-2 Z^2+\left| Z\right| \right) \notag\\
&+36 g_s M^2 N \pi  \left(17 (1+\squ) g_s M^2+16 (1-\squ) N \pi \right) \left(72+72 \squ+\pi ^2\right) \left| Z\right|^3 \notag\\
&+12 g_s M^2 N^2 \pi ^2 Z^2 \left(29 (1+\squ) \left(72+72 \squ+\pi ^2\right)+128 \left(-144 (1+\squ)+(2+3 \squ) \pi ^2\right) Z^2\right) \notag\\
&+144 g_s^2 M^4 N \pi  Z^4 \Big(3 \left(720+576 \squ-144 \squ^2-2 \pi ^2-11 \squ \pi ^2 -128 \left(24-\pi ^2\right) Z^2\right) \notag\\
&\hspace{3cm} -32 \left(72+\pi ^2+3 \squ \left(48-\pi ^2\right)\right) \left| Z\right| \Big) \notag\\
&+27 g_s^3 M^6 Z^4 \Big(13 (1+\squ) \left(72+72 \squ+\pi ^2\right)+64 \left(504-29 \pi ^2+6 \squ \left(-48+\pi ^2\right)\right) Z^2 \notag\\
&\hspace{3cm} -8 \left(-1584-1440 \squ+144 \squ^2+14 \pi ^2+29 \squ \pi ^2+384 \left(24-\pi ^2\right) Z^2\right) \left| Z\right| \Big) \Bigg]
\end{align}
Note that $s_1(Z)$, $s_2(Z)$ and $s_3(Z)$ are even functions of $Z$. Again, we expand the scalar field $Y(x^\mu, Z)$ with a set of complete eigenfunctions $\{ \Omega_n(Z), n\ge 1\}$ to separate its dependence on $x^\mu$ and $Z$.
\begin{align}
Y(x^{\mu}, Z) = \sum_{n=1}^{\infty} \mathcal{U}^{(n)}(x^\mu)\Omega_n(Z)
\end{align}
The eigenvalue equation and orthonormal condition that $\Omega_n$ must satisfy is determined by looking at eq. (\ref{EQ:SMDBIaction}). First of all, to fix the mass term, we must have the identity:
\begin{align}
T \int dZ \left( \, s_2(Z) \dot{\Omega}_m \dot{\Omega}_n + s_3(Z) \Omega_m \Omega_n \right)= \frac{1}{2} {m_n'}^2
\end{align}
We can obtain this condition by imposing the following eigenvalue equation and orthonormal condition:
\begin{align}
&T \int dZ \, s_1(Z) \, \Omega_m \Omega_n = \frac{1}{2}\delta_{m n} \nonumber\\
&- \partial_Z \left( s_2(Z) \partial_Z \Omega_n\right) + s_3(Z) \Omega_n = s_1(Z){m_n'}^2 \Omega_n \label{EQ:ScalarEigEq}
\end{align}
where ${m_n'}^2 \equiv \lambda_n' \mathcal{M}^2$ is the effective squared-mass of each scalar meson and $\lambda_n'$ is the eigenvalue of the corresponding mode. Consequently, we recover the action of a set of scalar fields with canonical kinetic terms:
\begin{align}
S_{\text{QCD, Scalar}} = - \int d^4x\, \sum_{n = 1}^{\infty} \left[ \frac{1}{2} \partial_\mu \mathcal{U}^{(n)} \partial^\mu\mathcal{U}^{(n)} + \frac{1}{2} \lambda_n' M^2 \left(\mathcal{U}^{(n)}\right)^2\right]
\end{align}

\section{Scalar Meson Spectrum \label{secmess}}

Similar to the case of vector mesons, we define a differential operator $\mathbf{H}_{\text{s}}$ from the eigenvalue equation (\ref{EQ:ScalarEigEq}) and consider perturbations with the controlling parameter $\delta = \frac{g_s M^2}{N}$. Thus, we can rewrite \eqref{EQ:ScalarEigEq} as:
\begin{align}
(\ref{EQ:ScalarEigEq}) &\rightarrow  - \frac{s_2(Z)}{\mathcal{M}^2 s_1(Z)} \left( \partial_Z^2 + \frac{s_2'(Z)}{s_2(Z)}  \, \partial_Z   - \frac{s_3(Z)}{s_2(Z)} \right) \Omega_n =  \lambda_n' \Omega_n
\end{align}
which can alternatively be expressed as an eigenvalue 
equation of the form:
\begin{equation}
\mathbf{H}_{\rm s} \vert\Omega_n\rangle = \lambda_n'\vert\Omega_n\rangle
\end{equation}
with the Hamiltonian now expressed as:
\begin{align}\label{hamilsca}
\mathbf{H}_{\text{s}} &\equiv   - \frac{s_2(Z)}{\mathcal{M}^2 s_1(Z)} \left( \partial_Z^2 + \frac{s_2'(Z)}{s_2(Z)}  \, \partial_Z   - \frac{s_3(Z)}{s_2(Z)} \right)
\end{align}
The orthogonality condition of the states $\Omega_n(Z) = \langle Z\vert \Omega_n\rangle$, or {alternatively} the ket-states
$\vert\Omega_n\rangle$, can be expressed succinctly using $G(Z) \equiv 2 \,T s_1(Z)$:
\begin{equation}\label{orthocon}
\langle \Omega_m | \Omega_n \rangle \equiv \int\limits_{-\infty}^{\infty} dZ \, G(Z) \Omega_m \Omega_n = \delta_{mn} 
\end{equation}
Having a glimpse at the future eigenfunctions that will be derived from the above eigenvalue equation, we decide to change the problem slightly. Instead of solving for the eigenfunctions $\Omega_n$, we postulate that:
\begin{equation} \label{ome}
\Omega_n = |Z| \omega_n
\end{equation}
and solve instead for $\omega_n$. Note that the eigenvalues of $\mathbf{h}_{\text{s}}$ are also eigenvalues of $\mathbf{H}_{\text{s}}$ and $\omega_n$ has the same $Z$ parity as $\Omega_n$. Performing this substitution at the level of the eigenvalue equation, we find a new differential operator $\mathbf{h}_{\text{s}}$ with $\omega_n$ as eigenfunctions and $\lambda_n'$ as eigenvalues.  The new eigenvalue problem becomes:
\begin{align}
\mathbf{h}_{\text{s}} | \omega_n \rangle &= \lambda_n' | \omega_n \rangle \label{EQ:Sqmeeq}
\end{align}
with the new Hamiltonian derived from \eqref{hamilsca}:
\begin{align}\label{hami}
\mathbf{h}_{\text{s}} &= - \frac{s_2(Z)}{M^2 s_1(Z)} \left[ \partial_Z^2 + \left(\frac{s_2'(Z)}{s_2(Z)} + \frac{2}{Z} \right) \, \partial_Z   - \frac{s_3(Z)}{s_2(Z)} + \frac{1}{Z}\frac{s_2'(Z)}{s_2(Z)} + 2\, \frac{\delta(Z)}{|Z|}\right] 
\end{align}
The wavefunctions $\omega_n(Z)$ and $\Omega_n(Z)$ are expectedly related by \eqref{ome},
which means that the orthogonality condition between the ket states $\vert\omega_n\rangle$ can again be expressed as \eqref{orthocon} except with $G(Z)$ replaced by $g(Z)$, where:
\begin{align}
g(Z) &= Z^2 G(Z)
\end{align}
For all matter and purposes, the delta function term is set to zero since we solve the differential piecewise. Therefore, the $1/|Z|$ singularity will not cause any trouble.

\subsection{Zeroth-Order Eigenvalues and Eigenfunctions \label{secmeso}}

The zeroth-order eigenvalue equation that we must first solve is summarized as follows:
\begin{align}
\mathbf{h}_{\text{s}}^{(0)} | \omega_n^{(0)} \rangle &= \lambda_n'^{(0)} | \omega_n^{(0)} \rangle \quad \text{with} \quad \langle \omega_m^{(0)} | \omega_n^{(0)} \rangle^{(0)} = \delta_{mn} \label{EQ:Snormcond}
\end{align}
where the zeroth-order Hamiltonian can be easily derived from \eqref{hami}:
\begin{align}
\mathbf{h}_{\text{s}}^{(0)} &=  - e^{2 |Z|} \left( \partial_Z^2 + 4 \, \text{sgn}(Z) \, \partial_Z \right) \end{align}
with the zeroth-order {orthogonality function} $g^{(0)}$ {given by}:
\begin{align}
g^{(0)} &=  T \frac{\left(1 + \squ + \frac{\pi^2}{24}\right)}{\sqrt{1+\squ}} \frac{27 \cA^{4/3} c^2 (3 \, g_s N^5 \pi^{11})^{1/4} }{2 \sqrt{2}} \, e^{2 |Z|}
\end{align}
As in the vector mesons' case, the differential operator $\mathbf{h}_{\text{s}}^{(0)}$ is invariant under $Z \rightarrow - Z$, which means that the solutions are either odd or even in $Z$. The zeroth-order eigenvalue equation is:
\begin{align}
\ddot{\omega}_n(Z)^{(0)}& + 4 \, \text{sgn}(Z) \dot{\omega}_n^{(0)}(Z)  + e^{- 2|Z|}  \lambda_n' \omega_n^{(0)}(Z) = 0 \label{EQ:Sde}
\end{align}

Solving the differential equation (\ref{EQ:Snormcond}) in the regimes $Z > 0$ and $Z < 0$, we obtain piecewise solutions given in terms of Bessel's functions as well.
\begin{align}
\omega_n^{(0)}(Z) = C_n' \, e^{-2|Z|}  J_2\left(\sqrt{\lambda_n'} e^{-|Z|}\right) + D_n' \, e^{-2|Z|}  Y_2\left(\sqrt{\lambda_n'} e^{-|Z|}\right)
\end{align}
Again, we set $D_n' = 0$ to have normalizable modes and $C_n'$ are determined by using eq.(\ref{EQ:Snormcond}) and the following manipulations:
\begin{align}
&\int_{-\infty}^{0^-} g^{(0)}(Z) (\omega_n^{(0)} (Z))^2 dZ + \int_{0^+}^{\infty} g^{(0)}(Z) (\omega_n^{(0)} (Z))^2 dZ = 1 \nonumber\\
&{g'}^{(0)} \left( \int_{-\infty}^{0^-} e^{2|Z|} (\omega_n^{(0)} (Z))^2 dZ + \int_{0^+}^{\infty} e^{2|Z|} (\omega_n^{(0)}(Z))^2 dZ \right) = 1 \nonumber\\
&{g'}^{(0)} {C_n'}^2 \left( \frac{1}{\lambda'_n}\right) \left(\lambda_n' {J_0}^2\left(\sqrt{\lambda_n'} \right) + (\lambda_n' - 4) {J_1}^2\left(\sqrt{\lambda_n'}\right) \right) = 1 
\end{align}
which gives us the functional form of $C_n'$:
\begin{align}
C'_n = \sqrt{\frac{\lambda_n'}{{g'}^{(0)}}} \frac{1}{\sqrt{\lambda_n' {J_0}^2\left(\sqrt{\lambda_n'} \right) + (\lambda_n' - 4) {J_1}^2\left(\sqrt{\lambda_n'}\right)}}
\end{align}
The eigenvalues are obtained by solving the following odd and even conditions, respectively. These conditions again guarantee perfect orthonormality of the eigenfunctions (see eq. \ref{EQ:SOrthCond}).
\begin{align}
\omega_n^{(0)}(0,\lambda_n') = J_2(\sqrt{\lambda_n'}) &= 0\quad (\text{Odd functions})\label{EQ:Soddcond}\\
\partial_Z \omega_n^{(0)}(0,\lambda_n') = J_1(\sqrt{\lambda_n'}) &= 0 \quad (\text{Even functions})\label{EQ:Sevencond}
\end{align}
The zeroes can also be found graphically by looking at {\bf figure \ref{zeross}}:
\begin{figure}[H]
\centering
\begin{subfigure}{0.48\textwidth}
\includegraphics[width=\textwidth]{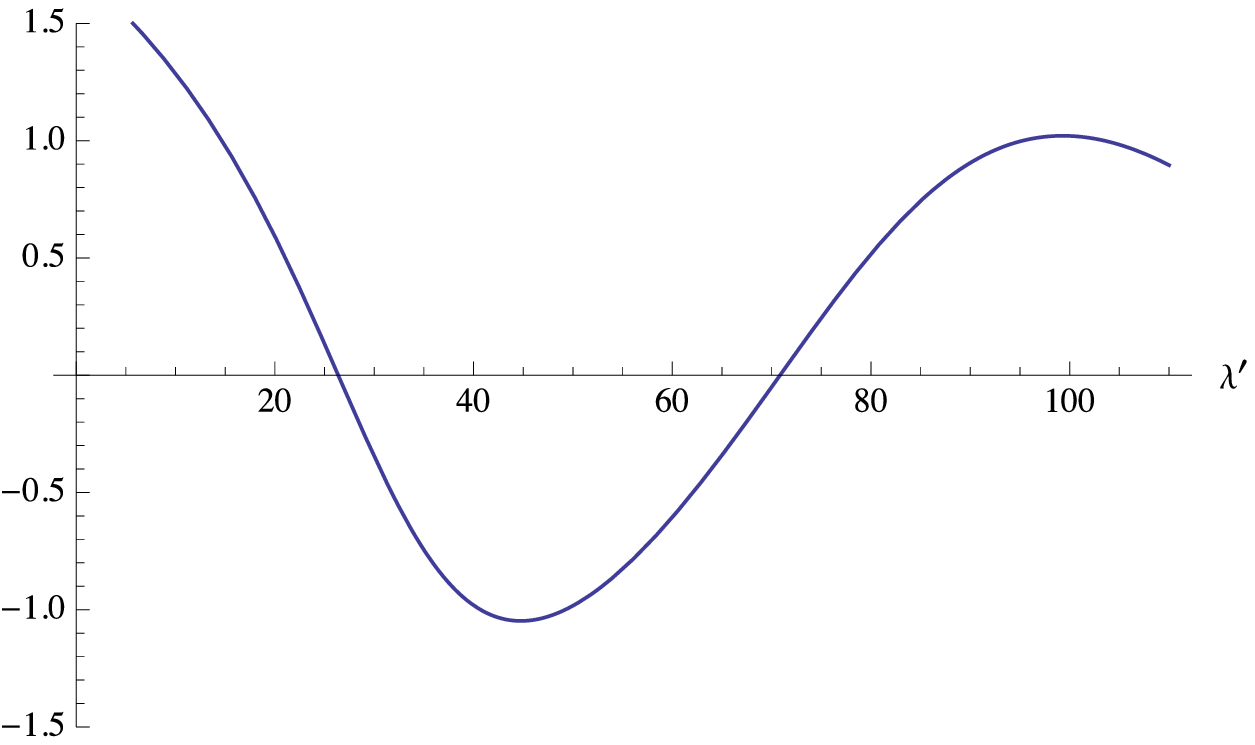}
\subcaption{$\omega_n(0, \lambda_n')$}
\end{subfigure}
\begin{subfigure}{0.48\textwidth}
\includegraphics[width=\textwidth]{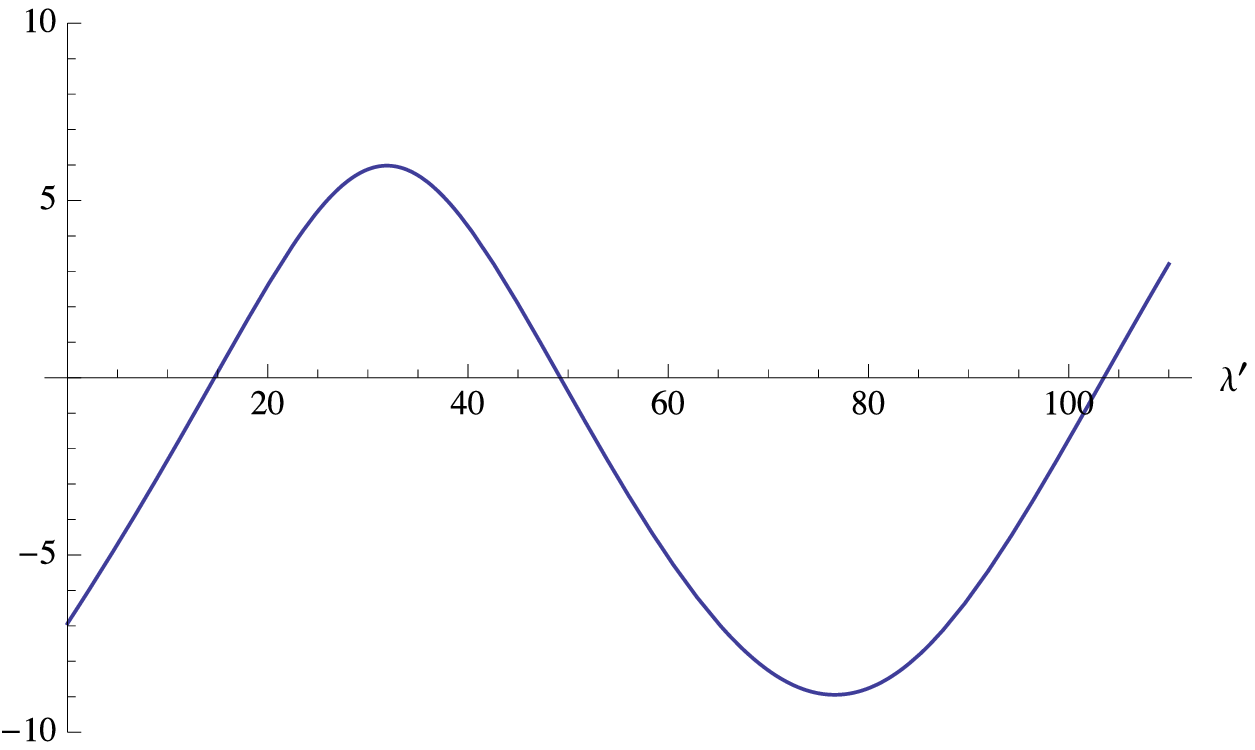}
\subcaption{$\partial_Z \omega_n(0,\lambda_n')$}
\end{subfigure}
\caption{Zeroes of $\omega_n(0, \lambda_n')$ and $\partial_Z \omega_n(0,\lambda_n')$.}
\label{zeross}
\end{figure}
For odd functions, we add an extra $\text{sign}(Z)$ as we did for the vector meson eigenfunctions.
\begin{align}
&\omega_{2n + 1}^{(0)}(Z) = C'_{2n+1} \, e^{-2|Z|}  J_2\left(\sqrt{\lambda_{2n+1}'} e^{-|Z|}\right) \nonumber\\
&\omega_{2n}^{(0)}(Z) = C'_{2n} \, \text{sgn}(Z) \, e^{-2|Z|}  J_2\left(\sqrt{\lambda_{2n}'} e^{-|Z|}\right)
\end{align}

\begin{figure}[H]
\begin{tabular}{|c|c|c|}
\hline
\begin{subfigure}[H]{0.3\textwidth}
\vspace{0.1 cm}
\includegraphics[width=\textwidth]{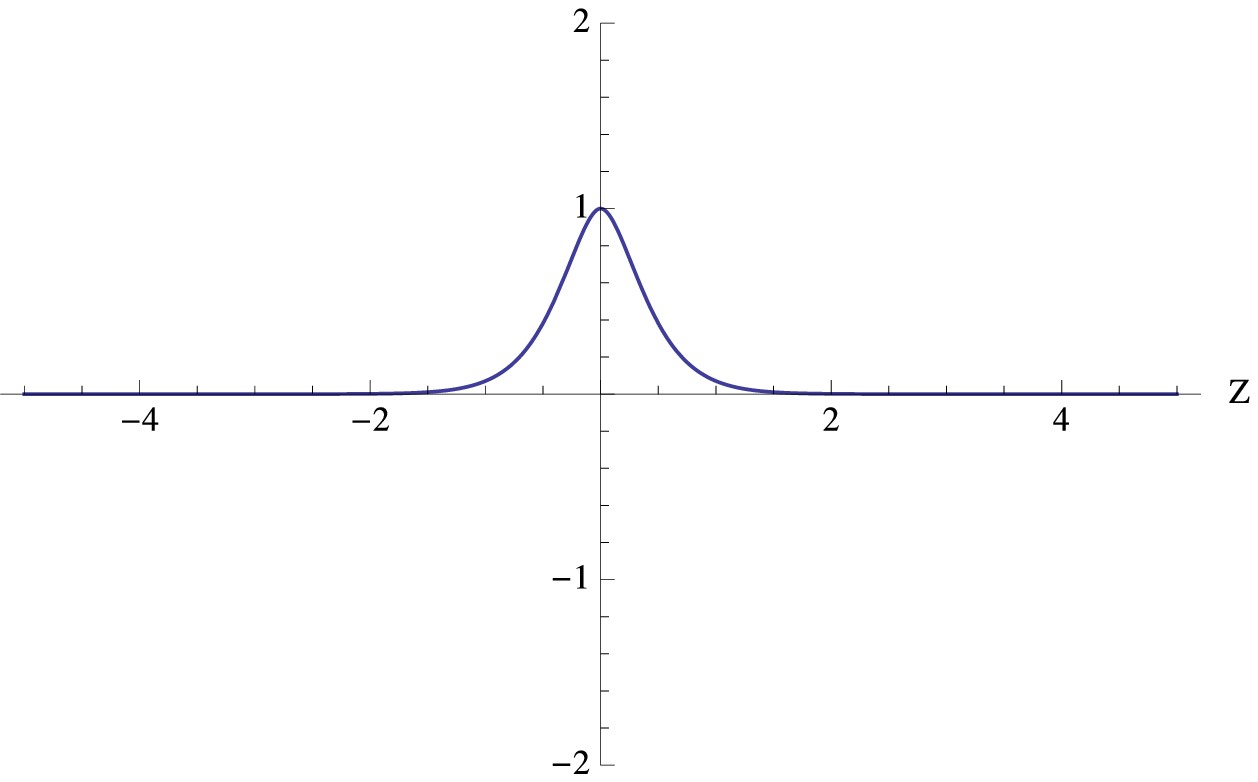}
\subcaption{$\omega_1(Z)$}
\end{subfigure}
&
\begin{subfigure}[H]{0.3\textwidth}
\vspace{0.1 cm}
\includegraphics[width=\textwidth]{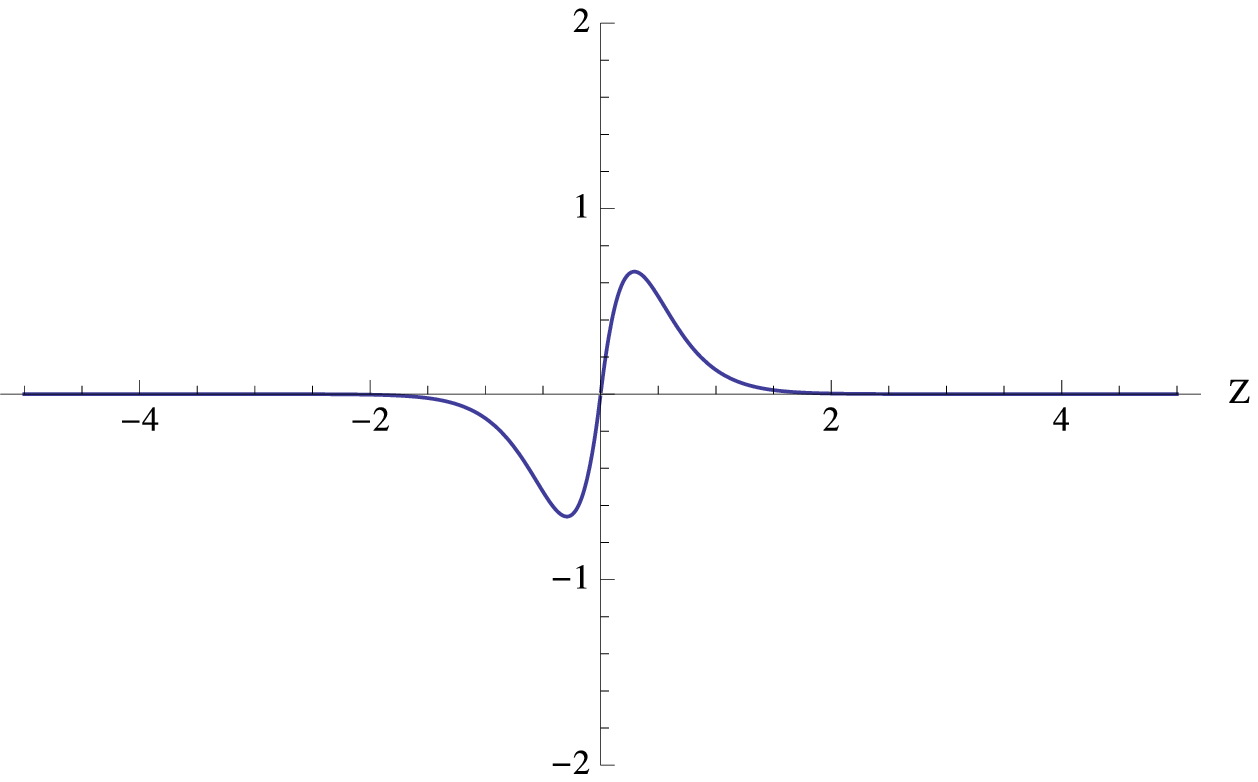}
\subcaption{$\omega_2(Z)$}
\end{subfigure} 
&
\begin{subfigure}[H]{0.3\textwidth}
\vspace{0.1 cm}
\includegraphics[width=\textwidth]{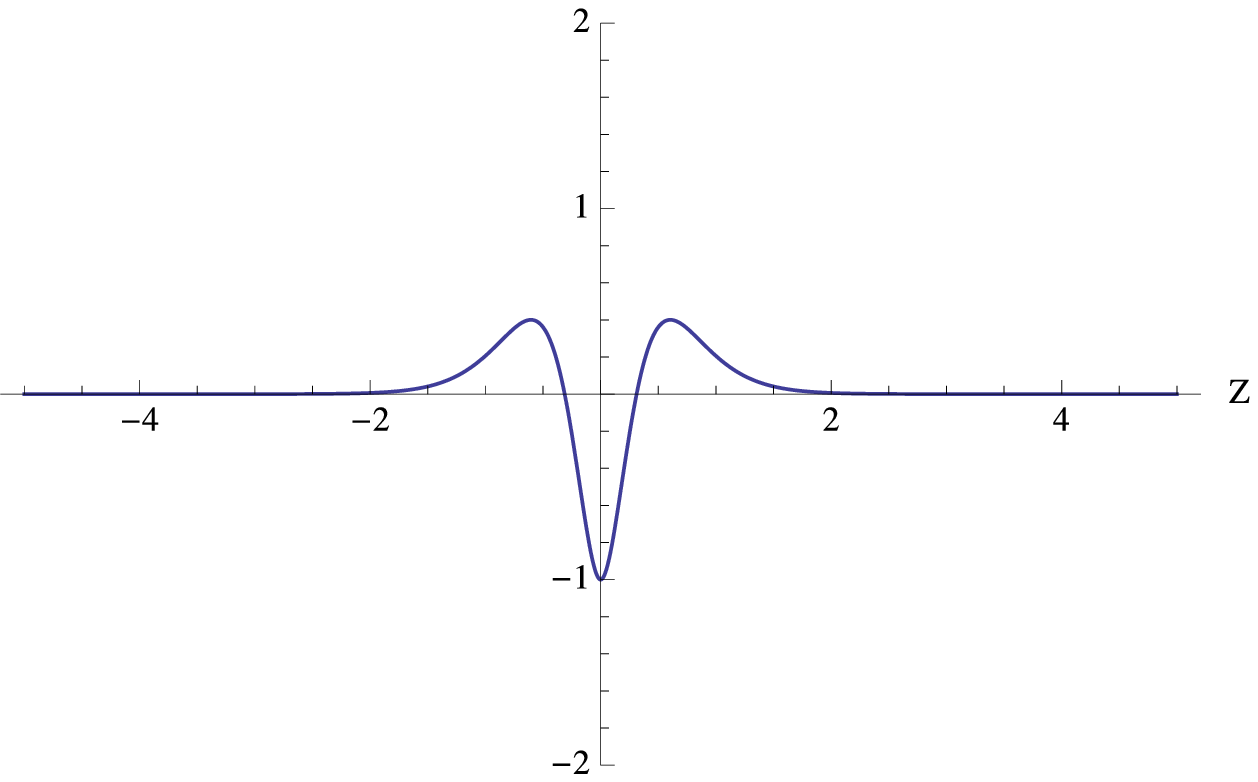}
\subcaption{$\omega_3(Z)$}
\end{subfigure}\\
\hline
\end{tabular}
\caption[Zeroth-order eigenfunctions of the three lightest scalar mesons in the MDGJ model.]{Zeroth-order eigenfunctions of the three lightest scalar mesons in the MDGJ model.}
\label{normalss}
\end{figure}

We also obtain perfect orthogonality between two different eigenfunctions. One can easily see this by evaluating the orthogonality integral between two odd (or even) functions and remembering eq. (\ref{EQ:Soddcond}) and (\ref{EQ:Sevencond}).
\begin{align}
\langle \omega_m^{(0)} | \omega_n^{(0)} \rangle^{(0)} &= \int\limits_{-\infty}^{\infty} dZ g^{(0)}(Z) \omega_m^{(0)}(Z) \omega_n^{(0)}(Z) \quad \text{$m \ne n$, $(m,n) \in 2 \mathbb{Z}$ or $2 \mathbb{Z} + 1$ }\nonumber\\
&= \frac{2 \, {g'}^{(0)} C'_m C'_n}{\lambda_m - \lambda_n} \left( \sqrt{\lambda_n}  \, J_1(\sqrt{\lambda_n})J_2(\sqrt{\lambda_m})- \sqrt{\lambda_m} \, J_1(\sqrt{\lambda_m})J_2(\sqrt{\lambda_n})\right) \label{EQ:SOrthCond}
\end{align}
Again, the orthogonality is no surprise since the differential equation (\ref{EQ:Sde}) can be cast into a Sturm-Liouville equation.

Now, let us comment on the ansatz of $\Omega_n$ that we used above. Had we instead pursue the analysis with $\Omega_n$, we would have unsurprisingly found $\Omega_n(Z) = C' \, |Z| e^{-2|Z|}  J_2\left(\sqrt{\lambda_n'} e^{-|Z|}\right)$. However, the odd-function condition would have been trivially satisfied and it would have given us no clue on the numerical values of the odd eigenvalues. We can also convince ourselves that the odd-function condition must be imposed on $\omega_n$ (as opposed to $\Omega_n$) since it is the only way to have perfect orthonormality.

\subsection{First-Order Eigenvalue \label{secmes1}}

We now look at the first-order correction to the eigenvalues of eq.(\ref{EQ:Sqmeeq}). The formula for such correction is given by the following expressions:
\begin{align}
\lambda_n'^{(1)} &= \langle \omega_n^{(0)}|\mathbf{h}_{\text{s}}^{(1)} | \omega_n^{(0)} \rangle^{(0)} 
\end{align}
with the first-order Hamiltonian now expressed as:
\begin{align}
\mathbf{h}_{\text{s}}^{(1)} &= \frac{3 \, e^{2 |Z|}}{2 \pi} \left[ |Z| \, \partial_Z^2 + \left( \frac{5}{4} \, \text{sgn}(Z) + 4 \, Z \left( \frac{(13 \squ+15)}{1+\squ}-\frac{288 (3 \squ + 4)}{72 (1 + \squ ) + \pi^2} \right) \right) \partial_Z \right]
\end{align}
When operating $\mathbf{h}_{\text{s}}^{(1)}$ on $\omega_n^{(0)}$, we identify $|Z|', |Z|''$ with $\text{sign}(Z),\, 2 \,\delta(Z)$ respectively. Again, the first-order correction depends on $\delta$ and $\squ$ and are assigned numerical values as explained in sec. \ref{S:Sum}.

\subsection{Meson Identifications \label{secmesid}}

To identify the scalar mesons with PDG 
values \cite{PDG},
first, we must find the parity eigenvalue ($\omega_{\mathcal{P},n} = \pm 1$) of the scalar mesons $\mathcal{U}^{(n)}$. Recall the action of the parity operator from eq. (\ref{EQ:ParityOp}). In terms of $\omega_n$, the $Y$ expansion is expressed as follows:
\begin{align}
Y(x^{\mu}, Z) = \sum_{n=1}^{\infty} \mathcal{U}^{(n)}(x^\mu)|Z|\omega_n(Z) \label{EQ:YExp}
\end{align}
Looking at eq. (\ref{EQ:YExp}), we conclude that $\mathcal{U}^{(n)}$  must be odd (resp. even) under parity when $\omega_n$ is odd (resp. even) in order for $Y(x^\mu,Z)$ to be a five-dimensional scalar.

Second, looking at the Chern-Simons coupling to determine the charge conjugation eigenvalue ($\omega_{\mathcal{C},n} = \pm 1$), we find that 
$Y$ is also even under C-parity. Then, $\mathcal{U}^{(n)}$ is even (resp. odd) under charge conjugation when $\omega_n(Z)$ is even (resp. odd). Knowing the eigenvalues of each scalar mesons under $\mathcal{P}$ and 
$\mathcal{C}$, we can try cross-referencing them with the Particle Data Group (PDG) database \cite{PDG} again by specifying to fields that are vectors of the approximate isospin $SU(2)$ symmetry. Although our model is predicting $J^{CP} = 0^{--}$ states, such particles aren't found in QCD. This comes from the fact that a $q \bar{q}$ bound state of quarks has a C-parity eigenvalue equal to $(-1)^{l+s}$ where $l$ and $s$ are the orbital and spin angular momentum respectively. Since scalar mesons have $\textbf{J} = \textbf{L} + \textbf{S} = 0$, either $l,s = 0$ or $l,s = 1$, both cases giving a positive C-parity eigenvalue. The following table summarizes our knowledge of the first three scalar mesons. 
\begin{table}[H]
\centering
\caption{Scalar Mesons}
\begin{tabular}{| c | c | c | c | c |} 
\hline
 & $\lambda_n'^{(0)}$ & $\omega_n^{(0)}$ Parity &$\omega_{\mathcal{C}}$ & $\omega_{\mathcal{P}}$ \\ \hline
$\mathcal{U}^{(1)}$ & 14.68 & Even & + & + \\ \hline
$\mathcal{U}^{(2)}$ & 26.34 & Odd  & - & - \\ \hline
$\mathcal{U}^{(3)}$ & 49.22 & Even & + & + \\ \hline
\end{tabular}
\end{table}
However, a more detailed analysis along the lines of \cite{SSrecent} shows that the scalar mesons that we study here are {\it not} related to the QCD scalar mesons. In fact, the QCD scalar mesons should be related to massive modes of the open string in the RR background of Region 1. The massive KK modes from the massless sectors are extra states not found in actual QCD.

\section{Mass ratio calculations \label{S:Sum}}

By taking the zeroth and first order eigenvalues into account, we can predict values of squared-mass ratios using this formula:
\begin{align}
R_{n/m} \equiv \frac{\lambda_n}{\lambda_m} = \frac{\lambda_n^{(0)} + \delta \, \lambda_n^{(1)} + O(\delta^2)}{\lambda_m^{(0)} + \delta \, \lambda_m^{(1)} + O(\delta^2)} \approx \frac{\lambda_n^{(0)}}{\lambda_m^{(0)}} + \frac{\delta}{\left(\lambda_m^{(0)}\right)^2} \left(  \lambda_m^{(0)}\lambda_n^{(1)} - \lambda_n^{(0)} \lambda_m^{(1)} \right) \label{EQ:Massratio}
\end{align}
We stop at first-order terms in the $\delta$ expansion, since our analysis hasn't considered the higher-order contributions. This formula is correct for both vector-to-vector and scalar-to-vector ratios hence $\lambda_n$ and $\lambda_m$ stand for generic mesons' eigenvalues.
\nl

In order to obtain numerical estimates for $R_{n/m}$, we determine the arbitrary parameter $\delta$ and $\squ$ in two ways. For each of these ways, we present a summary table of our predictions for $R_{n/1}$ compared with the estimates of the Sakai-Sugimoto model and the PDG value $R_{n/1}^{\text{PDG}}$. We took care to label our eigenvalues in the same way as in the Sakai-Sugimoto model to facilitate their comparison. The fourth column presents our zeroth-order estimates of the ratios, symbolized as $R_{n/m}^{(0)} \equiv \frac{\lambda_n^{(0)}}{\lambda_m^{(0)}}$, in order to see the effect of the first-order correction.

{Before moving on with the numerical estimates of the mass ratios, we point out a relevant caveat of this analysis. As eluded earlier, the contributions of the massive open string sector to the mesons' masses have been omitted. When studying the mesons' spectrum arising from the massive open string sector, one can ask an obvious question: Should the PDG values be compared with the massive KK modes (derived from the open string massless sector) or the massive open string modes? To answer this question, one needs to quantize the open string in the RR backgrounds of Regions 1, 2 and 3, as there is no reason for open strings to localize in Region 1 only. A large 'tHooft coupling simplified the analysis of \cite{SSrecent}, but it is not clear that such simplifications will carry through to our model as the RR backgrounds may not decouple even for large 'tHooft coupling. Therefore, the study of the massive open string modes requires a more sophisticated analysis, which has never been done to this day. Fortunately, this exercise is not necessary here since our predictions of the mesons' mass ratios fall within the range of the QCD mass ratios, which justifies the neglect of the massive open string modes.}

First, we determine $\delta$ and $\squ$ by minimizing the vector mesons' $\chi^2/\text{DOF}$. To calculate $\chi^2/\text{DOF}$, we use the following formula:
\begin{align}
\chi^2/\text{DOF} &= \frac{1}{2}\sum_{i=1}^4 \left( \frac{R_{i/1} - R_{i/1}^{\text{PDG}}}{\Delta R_{i/1}^{\text{PDG}}}\right)^2 \\
\Delta R_{i/j}^{\text{PDG}} &= 2 \left(\frac{M_i^{\text{PDG}}}{M_j^{\text{PDG}}}\right)^2 \sqrt{\left(\frac{\Gamma_i^{\text{PDG}}}{M_i^{\text{PDG}}}\right)^2 + \left(\frac{\Gamma_j^{\text{PDG}}}{M_j^{\text{PDG}}}\right)^2}
\end{align}
where $M_i^{\text{PDG}}$ and $\Gamma_i^{\text{PDG}}$ refer to the PDG value of the $i^{\text{th}}$ particle's mass and width, respectively. The $\chi^2/\text{DOF}$ allows us to compare the overall accuracy of the model with Sakai and Sugimoto's predictions, which have a $\chi^2/4 = 1.1874$.

Since we fix the numerical value of $\delta$ and $\squ$ \textit{a posteriori}, we have to make sure that our minimization process explores a parameter space that is consistent with the expansions that we have done above. The restrictions that we chose to impose on the parameter space are the following:
\begin{table}[H]
\centering
\label{EQ:MinCond1}
\caption{Mimization Process Conditions.}
\begin{tabular}{|cl|}
\hline
1. & $0 < \delta < 1$ \\ \hline
2. & $\squ > 0$  \\ \hline
3. & $|\delta \, \lambda_n^{(1)}| < 0.3 \, \lambda_n^{(0)}$  \text{(For all eigenvalues)} \\ \hline
\end{tabular}
\label{T:mincond}
\end{table}

The first condition guarantees that our $\delta$ expansions are valid. The second condition prohibits a negative squashing of one of the conifold sphere. The third condition makes sure that the expansion that is done in eq. (\ref{EQ:Massratio}) is valid.

The following table gives some numerical estimates of the vector meson mass ratios with the corresponding Sakai-Sugimoto results \cite{SS}.

\begin{table}[H]
\caption{Mass ratio calculations with $\delta = 0.8730$ and $\squ = 1.0181$ minimizing $\chi^2/2$ to $1.4068$.}
\begin{equation*}
\begin{array}{|l|c|c|c|c|c|}
\hline
 & \lambda_n/\lambda_m & \text{Sakai-Sugimoto} & R_{n/m}^{(0)} & R_{n/m} & \text{Exp. Value} \, R_{n/m}^{\text{PDG}} \\
\hline
 m_{a_1(1260)}^2/m_{\rho (770)}^2 & \lambda _2/\lambda _1 & 2.32 & 2.54 & 2.20 & 2.52 \\
\hline
 m_{\rho (1450)}^2/m_{\rho (770)}^2 & \lambda _3/\lambda _1 & 4.22 & 5.27 & 4.06 & 3.57 \\
\hline
 m_{a_1(1640)}^2/m_{\rho (770)}^2 & \lambda _4/\lambda _1 & 6.62 & 8.51 & 5.99 & 4.51 \\
\hline
 m_{\rho (1700)}^2/m_{\rho (770)}^2 & \lambda _5/\lambda _1 & 9.53 & 12.95 & 8.50 & 4.92 \\
\hline
\end{array}
\end{equation*}
\label{T:minchi2}
\end{table}
In the table above, we see that the results compare well with the Sakai-Sugimoto analysis \cite{SS} with the squashing factor given by:
\begin{equation}
h_4^a = 1 + {1.0181\over \sin^2\theta_2}
\end{equation}
Note however that the $1^{--}$ states $\rho(1450)$ and $\rho(1700)$ could also appear from the massive stringy sector of the model (see \cite{SSrecent} for a discussion on this). However, since we haven't been able to quantize open string states beyond the massless level here, we cannot verify the masses for these mesons {coming} from the next excited state. {We assume} that these states appear from the massive KK modes of the massless open string sector. The other three states, namely $1^{--}[\rho(770)]$,
$1^{++}[a_1(1260)]$ and $1^{++}[a_1(1640)]$, appear only from the massless open string sector. 

We can also determine $\delta$ and $\squ$ by fixing the first ratio $R_{2/1}$ to its experimental value as shown in the following table.

\begin{table}[H]
\caption{Mass ratio calculations determining $\delta = 0.0564$ and $\squ = 0.9365$ by fixing $R_{n/1}$. \\ \text{\hspace{1.5cm}} $\chi^2/2 = 7.0279$}
\begin{equation*}
\begin{array}{|l|c|c|c|c|c|}
\hline
 & \lambda_n/\lambda_m & \text{Sakai-Sugimoto} & R_{n/m}^{(0)} & R_{n/m} & \text{Exp. Value} \, R_{n/m}^{\text{PDG}} \\
\hline
 m_{a_1(1260)}^2/m_{\rho (770)}^2 & \lambda _2/\lambda _1 & 2.32 & 2.54 & 2.52 & 2.52 \\
\hline
 m_{\rho (1450)}^2/m_{\rho (770)}^2 & \lambda _3/\lambda _1 & 4.22 & 5.27 & 5.19 & 3.57 \\
\hline
 m_{a_1(1640)}^2/m_{\rho (770)}^2 & \lambda _4/\lambda _1 & 6.62 & 8.51 & 8.34 & 4.51 \\
\hline
 m_{\rho (1700)}^2/m_{\rho (770)}^2 & \lambda _5/\lambda _1 & 9.53 & 12.95 & 12.65 & 4.92 \\
\hline
\end{array}
\end{equation*}
\end{table}
The squashing factor decreases a bit, but the overall ratios are not better than the ones in our earlier table. To see whether higher orders in the $\theta_2$ expansion have any effects on the mass ratio, we turn to the following analysis.

\subsection{Comments on the sixth-order $\theta_2$ expansion \label{sec6}}

We now assess the effect of considering higher orders in the $\theta_2$ expansion that was used in the vector and scalar meson actions (eq. \ref{EQ:VMDBIaction} \& \ref{EQ:SMDBIaction}). Since the mathematical expressions become very large in such case, we will not show all the derivations that we described above, but simply compare the mass ratios.
\nl
First of all, at $\delta = 0.8730$ and $\squ = 1.0181$ where the vector mesons $\chi^2/\text{DOF}$ derived from the second-order $\theta_2$ expansion is minimal, the sixth-order $\theta_2$ expansion slightly improves the overall ratios. However, this point breaks the third condition that we imposed earlier.

\begin{table}[H]
\caption{Comparison of mass ratios at different orders of the $\theta_2$ expansion.}
\begin{equation*}
\begin{array}{|c|c|c|c|c|}
\hline
 & \lambda_n/\lambda_m & R_{n/m} \left( O\left({\theta_2}^2\right)\right)& R_{n/m} \left( O\left({\theta_2}^6\right)\right)& \text{Exp. Value} \, R_{n/m}^{\text{PDG}} \\
\hline
 m_{a_1(1260)}^2/m_{\rho (770)}^2 & \lambda _2/\lambda _1 & 2.20 & 2.19 & 2.52 \\
\hline
 m_{\rho (1450)}^2/m_{\rho (770)}^2 & \lambda _3/\lambda _1 & 4.06 & 3.94 & 3.57 \\
\hline
 m_{a_1(1640)}^2/m_{\rho (770)}^2 & \lambda _4/\lambda _1 & 5.99 & 5.71 & 4.51 \\
\hline
 m_{\rho (1700)}^2/m_{\rho (770)}^2 & \lambda _5/\lambda _1 & 8.50 & 7.94 & 4.92 \\
\hline
\chi^2/ \text{DOF} && 1.4068 & 0.9943 & \\ 
\hline
\end{array}
\end{equation*}
\end{table}

Second, we tried to minimize the $\chi^2/\text{DOF}$ with the sixth-order $\theta_2$ expansion, while still allowing a $30 \%$ maximal correction for the eigenvalues. However, this minimum is attained at the right boundary of $\delta$, hence we reduced the maximal allowable value for $\delta$ to $0.7$. The resulting ratios are shown below.

\begin{table}[H]
\caption{$O\left({\theta_2}^6\right)$ mass ratio calculations with $0 < \delta < 1$ and a maximal correction of 30\%. \\ \text{\hspace{1.5cm}} $\delta$ = 1.0000 and $\squ$ = 1.7280 minimize $\chi^2/2$ to 1.0590.}
\begin{equation*}
\begin{array}{|c|c|c|c|c|c|}
\hline
 \text{} & \lambda _n/\lambda _m & \text{Sakai-Sugimoto} & R_{n/m}^{(0)} & R_{n/m} & \text{Exp. Value }R_{n/m}^{\text{PDG}} \\
\hline
 m_{a_1(1260)}^2/m_{\rho (770)}^2 & \lambda _2/\lambda _1 & 2.32 & 2.54 & 2.15 & 2.52 \\
\hline
 m_{\rho (1450)}^2/m_{\rho (770)}^2 & \lambda _3/\lambda _1 & 4.22 & 5.27 & 3.93 & 3.57 \\
\hline
 m_{a_1(1640)}^2/m_{\rho (770)}^2 & \lambda _4/\lambda _1 & 6.62 & 8.51 & 5.72 & 4.51 \\
\hline
 m_{\rho (1700)}^2/m_{\rho (770)}^2 & \lambda _5/\lambda _1 & 9.53 & 12.95 & 8.05 & 4.92 \\
\hline
\end{array}
\end{equation*}
\end{table}

\begin{table}[H]
\caption{$O\left({\theta_2}^6\right)$ mass ratio calculations with $0 < \delta < 0.7$ and a maximal correction of 30\%. \\ \text{\hspace{1.5cm}} $\delta$ = 0.7000 and $\squ$ = 1.1996 minimize $\chi^2/2$ to 1.9707.}
\begin{equation*}
\begin{array}{|c|c|c|c|c|c|}
\hline
 \text{} & \lambda _n/\lambda _m & \text{Sakai-Sugimoto} & R_{n/m}^{(0)} & R_{n/m} & \text{Exp. Value }R_{n/m}^{\text{PDG}} \\
\hline
 m_{a_1(1260)}^2/m_{\rho (770)}^2 & \lambda _2/\lambda _1 & 2.32 & 2.54 & 2.26 & 2.52 \\
\hline
 m_{\rho (1450)}^2/m_{\rho (770)}^2 & \lambda _3/\lambda _1 & 4.22 & 5.27 & 4.25 & 3.57 \\
\hline
 m_{a_1(1640)}^2/m_{\rho (770)}^2 & \lambda _4/\lambda _1 & 6.62 & 8.51 & 6.36 & 4.51 \\
\hline
 m_{\rho (1700)}^2/m_{\rho (770)}^2 & \lambda _5/\lambda _1 & 9.53 & 12.95 & 9.13 & 4.92 \\
\hline
\end{array}
\end{equation*}
\end{table}

The results improve mildly over the ones from Sakai-Sugimoto, but are not in very good agreement with the PDG values \cite{PDG}. {One solution to resolve this discrepancy would be to calculate the massive KK modes $m^{(k)}_{KK}$ and define the total meson mass in the following way}:
\begin{equation}
m^{(k)}_v \equiv m^{(k)}_{KK} + m_s^{(k)}
\end{equation}
where the superscript $k$ {enumerates the vector mesons.} But as we pointed out earlier, determining $m_s^{(k)}$ is technically challenging. Another {solution} would to be explore the landscape of flux vacua \cite{DRS} where the details of the background {($g_s, N, M, \squ$, etc)} would change, hoping to find a point where the vector meson spectrum is considerably improved. We {develop} {this solution} in the following section.

\subsection{Landscape and other values of $\delta$ and $\squ$ \label{landscape}}

In the previous two tables, where we predicted the mass ratios of various vector and scalar mesons, we see that the squashing factors \eqref{squaf} of the type IIB internal ($\theta_2, \phi_2$) spheres in \eqref{sukra} take the following values:
\begin{equation}\label{holval}
h_4^{b} = 1 + {1.7280 \over \sin^2\theta_2}, ~~~~~~~~
h_4^c = 1 + {1.1996\over \sin^2\theta_2}
\end{equation}
For fixed values of $g_s$, these two squashing choices should be thought of as two different points in the landscape of flux vacua for two different values of ($M, N$). Both these points are dual to gauge theories with 
different matter contents and different set of 
operators. The predictions for the vector mesonic spectrum are not always comparably better than the results of Sakai-Sugimoto \cite{SS}, however we can go to different points in the flux landscape where the vector mesonic spectrum is somewhat better than the existing results in the literature \cite{SS, other0, SS2, other1, 
other2}\footnote{See footnote \ref{improve} for a discussion on the {\it improvement}.}. These points in the landscape will determine different values for the squashing factor $h_4$ (or $\squ$). 

In the following, we present other methods that we can try to determine the numerical values of $\delta$ and $\squ$.

\vskip.1in

$\bullet$ We minimize the vector mesons $\chi^2/\text{DOF}$ by using the unexpanded form of the mass ratios (eq. \ref{EQ:Massratio}), which is:
\begin{align}
R_{n/m} \equiv \frac{\lambda_n}{\lambda_m} = \frac{\lambda_n^{(0)} + \delta \, \lambda_n^{(1)}}{\lambda_m^{(0)} + \delta \, \lambda_m^{(1)}} \label{EQ:Massratio2}
\end{align}
while using the same minimization conditions (see table \ref{T:mincond}). We find $\delta = 0.8730, \squ = 1.0181$ and the $\chi^2/\text{DOF}$ is minimized to $1.4068$. Note that these values are identical to the ones that we found in table \ref{T:minchi2}, which confirms that our mass ratio expansions were correct up to first-order terms in $\delta$. The type IIB squashing factor then becomes:
\begin{equation}
h_4^d = 1 + {1.0181\over \sin^2\theta_2}
\end{equation}

\vskip.1in

$\bullet$ In the minimization process, we can also modify the conditions that we imposed on $\delta$ and $\squ$ (table \ref{T:mincond}). We allow various boundary values for $\delta$ (from $[0,0.3]$ to $[0,0.5]$) and different maximal corrections (from $30 \%$ to $70 \%$). For each of these methods, we find the values of $\delta$ and $\squ$ by minimizing the vector mesons $\chi^2/\text{DOF}$. {The following tables present the two extreme cases and we refer the reader to Appendix \ref{land} for the complete list. }

\begin{table}[H]
\caption{Mass ratio calculations with $0 < \delta < 0.3$ and a maximal correction of 30\%. \\ \text{\hspace{1.5cm}} $\delta$ = 0.3000 and $\squ$ = 0.3180 minimize $\chi^2/2$ to 3.6738.}
\begin{equation*}
\begin{array}{|c|c|c|c|c|c|}
\hline
 \text{} & \lambda _n/\lambda _m & \text{Sakai-Sugimoto} & R_{n/m}^{(0)} & R_{n/m} & \text{Exp. Value }R_{n/m}^{\text{PDG}} \\
\hline
 m_{a_1(1260)}^2/m_{\rho (770)}^2 & \lambda _2/\lambda _1 & 2.32 & 2.54 & 2.41 & 2.52 \\
\hline
 m_{\rho (1450)}^2/m_{\rho (770)}^2 & \lambda _3/\lambda _1 & 4.22 & 5.27 & 4.67 & 3.57 \\
\hline
 m_{a_1(1640)}^2/m_{\rho (770)}^2 & \lambda _4/\lambda _1 & 6.62 & 8.51 & 7.22 & 4.51 \\
\hline
 m_{\rho (1700)}^2/m_{\rho (770)}^2 & \lambda _5/\lambda _1 & 9.53 & 12.95 & 10.56 & 4.92 \\
\hline
\end{array}
\end{equation*}
\end{table}

\begin{table}[H]
\caption{Mass ratio calculations with $0 < \delta < 0.5$ and a maximal correction of 70\%. \\ \text{\hspace{1.5cm}} $\delta$ = 0.5000 and $\squ$ = 0.1501 minimize $\chi^2/2$ to 1.1329.}
\begin{equation*}
\begin{array}{|c|c|c|c|c|c|}
\hline
 \text{} & \lambda _n/\lambda _m & \text{Sakai-Sugimoto} & R_{n/m}^{(0)} & R_{n/m} & \text{Exp. Value }R_{n/m}^{\text{PDG}} \\
\hline
 m_{a_1(1260)}^2/m_{\rho (770)}^2 & \lambda _2/\lambda _1 & 2.32 & 2.54 & 2.31 & 2.52 \\
\hline
 m_{\rho (1450)}^2/m_{\rho (770)}^2 & \lambda _3/\lambda _1 & 4.22 & 5.27 & 4.09 & 3.57 \\
\hline
 m_{a_1(1640)}^2/m_{\rho (770)}^2 & \lambda _4/\lambda _1 & 6.62 & 8.51 & 5.93 & 4.51 \\
\hline
 m_{\rho (1700)}^2/m_{\rho (770)}^2 & \lambda _5/\lambda _1 & 9.53 & 12.95 & 8.10 & 4.92 \\
\hline
\end{array}
\end{equation*}
\end{table}

For these cases, the type IIB squashing factor becomes:
\begin{equation}
h_4^e = 1 + {0.3180\over\sin^2\theta_2}, ~~~~~~
h_4^f = 1 + {0.1501\over\sin^2\theta_2}
\end{equation}
Clearly, there are various other points in the landscape parametrized not only by ($\delta, \squ$) but also by the maxiaml percentage of correction that we allow. In the original brane picture (both in type IIA and type IIB), these are parametrized by ($M, N$) and the value of the fluxes\footnote{The claim that there is some improvement in the tables presented in
Appendix \ref{land} should be viewed in the 
following sense: given {\it any} experimental value, we will be able to choose
appropriate values for $\delta, u$ and the type IIB background fluxes to
match the experimental value to a good extent at least for the mesons that appear from the massive KK states of the massless closed string sector. It is interesting to note that the present experimental
values for the mesons (that we studied in this paper) can be matched exactly
to the decimal orders if we consider the background associated with values of $\delta > 0.5$. This is a bit of a surprise because it is exactly the regime where
${\cal O}(\delta^2)$ and $u$ corrections plus the massive open string sector
become relevant. Additionally when we push the value of $a$ close to an uncontrollable regime ($a \sim 1$), we can get better results (i.e. the $\chi^2$ improves). However, the errors also become unbounded as demonstrated in Appendix \ref{landb}.
Thus there is some improvement, but any matching can only be performed completely once we have the full closed and open string massive modes in this background. This will require the study of quantization of strings in RR as well as curved backgrounds. \label{improve}}. 

\section{Conclusion and discussion \label{seccono}}

In this work we determined the scalar and the vector mesonic spectra of our UV complete model and compared our results with the ones obtained by Sakai-Sugimoto \cite{SS}. The outcome of our calculations depend on {our} choices of $\delta \equiv {g_sM^2\over N}$ and the squashing parameter $\squ$ that relates how the ($\theta_2, \phi_2$) sphere is squashed (both in the type IIB and the type IIA string theory). For certain values of ($\delta, \squ$) parametrizing the space of {the} flux landscape, the vector mesonic spectrum can be improved somewhat over the existing values in the literature. 

Our type IIA dual shows how certain other improvements can be made over the original Sakai-Sugimoto model, under which the high temperature limit can be easily studied without encountering the issues pointed out in \cite{manmor}. The mapping of the spectra to the PDG values \cite{PDG} can be done for the vector mesons, but not for the scalar mesons due to certain subtleties associated with an underlying $Z_2$ symmetry \cite{SSrecent}. In fact, our model {focuses} only {on} the massive KK modes associated with the {\it massless} open string sector. There exists towers of {\it massive} open string
modes that we haven't analyzed here because quantization of open strings in Region 1 itself is highly non-trivial. Fortunately however, many of the vector mesons that we study here using the massless open string sector match with the actual QCD mesons.

Our results are expressed in terms of ($N, M, \delta, \squ$) and every choice of this set specifies a point in the landscape of flux vacua \cite{DRS} (see also \cite{aalok} for a slightly different parametrization of vacua in our set-up). We observe that there exists points in the landscape where the spectrum can be better. These points are specified by different squashings of the resolved sphere in the resolved warped deformed conifold background {and} different choices of fluxes. On the dual gauge theory side at every such points of the landscape, only the IR physics becomes relevant for the computation of both vector and scalar mesonic spectra. It would be interesting to see how {the mass ratios change} once {the} massive open string modes are incorporated. 

\centerline{\bf Acknowledgements}

{We are especially grateful to} Shigeki Sugimoto for {his patience and detailed} explanations of the massive open string sector of the Sakai-Sugimoto model. We would also like to thank Peter Ouyang and Martin Kruczenski for helpful discussions, and Long Chen for help with some of the 
brane-construction figures. 
The work of K. D, C. G, and M. R is supported in part by the National Science and Engineering Research Council (NSERC). The work of O. T is supported in part by the 
Fonds de recherche du Quebec $-$ Nature et technologies (FRQNT)
grant. The work of M. M is supported in part by a DOE grant number DE-SC0007884. 

\newpage

\begin{appendices}

\section{More details on the Landscape of spectra \label{land}}

In section \ref{landscape}, we discussed other points in the landscape of flux vacua \cite{DRS} parametrized by ($N, M, \delta, u$) where the vector and scalar mesonic spectra could be studied. We also compared the massive vector KK modes with the PDG \cite{PDG} and the Sakai-Sugimoto \cite{SS, SS2} results. In the following, {we present our predictions for other points of the landscape}.

\begin{table}[H]
\caption{Mass ratio calculations with $0 < \delta < 0.4$ and a maximal correction of 30\%. \\ \text{\hspace{1.5cm}} $\delta$ = 0.4000 and $\squ$ = 0.4826 minimize $\chi^2/2$ to 3.1958.}
\begin{equation*}
\begin{array}{|c|c|c|c|c|c|}
\hline
 \text{} & \lambda _n/\lambda _m & \text{Sakai-Sugimoto} & R_{n/m}^{(0)} & R_{n/m} & \text{Exp. Value }R_{n/m}^{\text{PDG}} \\
\hline
 m_{a_1(1260)}^2/m_{\rho (770)}^2 & \lambda _2/\lambda _1 & 2.32 & 2.54 & 2.37 & 2.52 \\
\hline
 m_{\rho (1450)}^2/m_{\rho (770)}^2 & \lambda _3/\lambda _1 & 4.22 & 5.27 & 4.57 & 3.57 \\
\hline
 m_{a_1(1640)}^2/m_{\rho (770)}^2 & \lambda _4/\lambda _1 & 6.62 & 8.51 & 7.01 & 4.51 \\
\hline
 m_{\rho (1700)}^2/m_{\rho (770)}^2 & \lambda _5/\lambda _1 & 9.53 & 12.95 & 10.20 & 4.92 \\
\hline
\end{array}
\end{equation*}
\end{table}

\begin{table}[H]
\caption{Mass ratio calculations with $0 < \delta < 0.5$ and a maximal correction of 30\%. \\ \text{\hspace{1.5cm}} $\delta$ = 0.5000 and $\squ$ = 0.6233 minimize $\chi^2/2$ to 2.7526.}
\begin{equation*}
\begin{array}{|c|c|c|c|c|c|}
\hline
 \text{} & \lambda _n/\lambda _m & \text{Sakai-Sugimoto} & R_{n/m}^{(0)} & R_{n/m} & \text{Exp. Value }R_{n/m}^{\text{PDG}} \\
\hline
 m_{a_1(1260)}^2/m_{\rho (770)}^2 & \lambda _2/\lambda _1 & 2.32 & 2.54 & 2.34 & 2.52 \\
\hline
 m_{\rho (1450)}^2/m_{\rho (770)}^2 & \lambda _3/\lambda _1 & 4.22 & 5.27 & 4.46 & 3.57 \\
\hline
 m_{a_1(1640)}^2/m_{\rho (770)}^2 & \lambda _4/\lambda _1 & 6.62 & 8.51 & 6.79 & 4.51 \\
\hline
 m_{\rho (1700)}^2/m_{\rho (770)}^2 & \lambda _5/\lambda _1 & 9.53 & 12.95 & 9.84 & 4.92 \\
\hline
\end{array}
\end{equation*}
\end{table}

\begin{table}[H]
\caption{Mass ratio calculations with $0 < \delta < 0.3$ and a maximal correction of 40\%. \\ \text{\hspace{1.5cm}} $\delta$ = 0.3000 and $\squ$ = 0.1728 minimize $\chi^2/2$ to 3.1237.}
\begin{equation*}
\begin{array}{|c|c|c|c|c|c|}
\hline
 \text{} & \lambda _n/\lambda _m & \text{Sakai-Sugimoto} & R_{n/m}^{(0)} & R_{n/m} & \text{Exp. Value }R_{n/m}^{\text{PDG}} \\
\hline
 m_{a_1(1260)}^2/m_{\rho (770)}^2 & \lambda _2/\lambda _1 & 2.32 & 2.54 & 2.40 & 2.52 \\
\hline
 m_{\rho (1450)}^2/m_{\rho (770)}^2 & \lambda _3/\lambda _1 & 4.22 & 5.27 & 4.58 & 3.57 \\
\hline
 m_{a_1(1640)}^2/m_{\rho (770)}^2 & \lambda _4/\lambda _1 & 6.62 & 8.51 & 7.00 & 4.51 \\
\hline
 m_{\rho (1700)}^2/m_{\rho (770)}^2 & \lambda _5/\lambda _1 & 9.53 & 12.95 & 10.13 & 4.92 \\
\hline
\end{array}
\end{equation*}
\end{table}

\begin{table}[H]
\caption{Mass ratio calculations with $0 < \delta < 0.4$ and a maximal correction of 40\%. \\ \text{\hspace{1.5cm}} $\delta$ = 0.4000 and $\squ$ = 0.3180 minimize $\chi^2/2$ to 2.6847.}
\begin{equation*}
\begin{array}{|c|c|c|c|c|c|}
\hline
 \text{} & \lambda _n/\lambda _m & \text{Sakai-Sugimoto} & R_{n/m}^{(0)} & R_{n/m} & \text{Exp. Value }R_{n/m}^{\text{PDG}} \\
\hline
 m_{a_1(1260)}^2/m_{\rho (770)}^2 & \lambda _2/\lambda _1 & 2.32 & 2.54 & 2.37 & 2.52 \\
\hline
 m_{\rho (1450)}^2/m_{\rho (770)}^2 & \lambda _3/\lambda _1 & 4.22 & 5.27 & 4.47 & 3.57 \\
\hline
 m_{a_1(1640)}^2/m_{\rho (770)}^2 & \lambda _4/\lambda _1 & 6.62 & 8.51 & 6.79 & 4.51 \\
\hline
 m_{\rho (1700)}^2/m_{\rho (770)}^2 & \lambda _5/\lambda _1 & 9.53 & 12.95 & 9.77 & 4.92 \\
\hline
\end{array}
\end{equation*}
\end{table}

\begin{table}[H]
\caption{Mass ratio calculations with $0 < \delta < 0.5$ and a maximal correction of 40\%. \\ \text{\hspace{1.5cm}} $\delta$ = 0.5000 and $\squ$ = 0.4440 minimize $\chi^2/2$ to 2.2805.}
\begin{equation*}
\begin{array}{|c|c|c|c|c|c|}
\hline
 \text{} & \lambda _n/\lambda _m & \text{Sakai-Sugimoto} & R_{n/m}^{(0)} & R_{n/m} & \text{Exp. Value }R_{n/m}^{\text{PDG}} \\
\hline
 m_{a_1(1260)}^2/m_{\rho (770)}^2 & \lambda _2/\lambda _1 & 2.32 & 2.54 & 2.33 & 2.52 \\
\hline
 m_{\rho (1450)}^2/m_{\rho (770)}^2 & \lambda _3/\lambda _1 & 4.22 & 5.27 & 4.37 & 3.57 \\
\hline
 m_{a_1(1640)}^2/m_{\rho (770)}^2 & \lambda _4/\lambda _1 & 6.62 & 8.51 & 6.57 & 4.51 \\
\hline
 m_{\rho (1700)}^2/m_{\rho (770)}^2 & \lambda _5/\lambda _1 & 9.53 & 12.95 & 9.41 & 4.92 \\
\hline
\end{array}
\end{equation*}
\end{table}

\begin{table}[H]
\caption{Mass ratio calculations with $0 < \delta < 0.3$ and a maximal correction of 50\%. \\ \text{\hspace{1.5cm}} $\delta$ = 0.3000 and $\squ$ = 0.0731 minimize $\chi^2/2$ to 2.6184.}
\begin{equation*}
\begin{array}{|c|c|c|c|c|c|}
\hline
 \text{} & \lambda _n/\lambda _m & \text{Sakai-Sugimoto} & R_{n/m}^{(0)} & R_{n/m} & \text{Exp. Value }R_{n/m}^{\text{PDG}} \\
\hline
 m_{a_1(1260)}^2/m_{\rho (770)}^2 & \lambda _2/\lambda _1 & 2.32 & 2.54 & 2.39 & 2.52 \\
\hline
 m_{\rho (1450)}^2/m_{\rho (770)}^2 & \lambda _3/\lambda _1 & 4.22 & 5.27 & 4.48 & 3.57 \\
\hline
 m_{a_1(1640)}^2/m_{\rho (770)}^2 & \lambda _4/\lambda _1 & 6.62 & 8.51 & 6.79 & 4.51 \\
\hline
 m_{\rho (1700)}^2/m_{\rho (770)}^2 & \lambda _5/\lambda _1 & 9.53 & 12.95 & 9.69 & 4.92 \\
\hline
\end{array}
\end{equation*}
\end{table}

\begin{table}[H]
\caption{Mass ratio calculations with $0 < \delta < 0.4$ and a maximal correction of 50\%. \\ \text{\hspace{1.5cm}} $\delta$ = 0.4000 and $\squ$ = 0.2037 minimize $\chi^2/2$ to 2.2184.}
\begin{equation*}
\begin{array}{|c|c|c|c|c|c|}
\hline
 \text{} & \lambda _n/\lambda _m & \text{Sakai-Sugimoto} & R_{n/m}^{(0)} & R_{n/m} & \text{Exp. Value }R_{n/m}^{\text{PDG}} \\
\hline
 m_{a_1(1260)}^2/m_{\rho (770)}^2 & \lambda _2/\lambda _1 & 2.32 & 2.54 & 2.36 & 2.52 \\
\hline
 m_{\rho (1450)}^2/m_{\rho (770)}^2 & \lambda _3/\lambda _1 & 4.22 & 5.27 & 4.38 & 3.57 \\
\hline
 m_{a_1(1640)}^2/m_{\rho (770)}^2 & \lambda _4/\lambda _1 & 6.62 & 8.51 & 6.57 & 4.51 \\
\hline
 m_{\rho (1700)}^2/m_{\rho (770)}^2 & \lambda _5/\lambda _1 & 9.53 & 12.95 & 9.33 & 4.92 \\
\hline
\end{array}
\end{equation*}
\end{table}

\begin{table}[H]
\caption{Mass ratio calculations with $0 < \delta < 0.5$ and a maximal correction of 50\%. \\ \text{\hspace{1.5cm}} $\delta$ = 0.5000 and $\squ$ = 0.3180 minimize $\chi^2/2$ to 1.8532.}
\begin{equation*}
\begin{array}{|c|c|c|c|c|c|}
\hline
 \text{} & \lambda _n/\lambda _m & \text{Sakai-Sugimoto} & R_{n/m}^{(0)} & R_{n/m} & \text{Exp. Value }R_{n/m}^{\text{PDG}} \\
\hline
 m_{a_1(1260)}^2/m_{\rho (770)}^2 & \lambda _2/\lambda _1 & 2.32 & 2.54 & 2.32 & 2.52 \\
\hline
 m_{\rho (1450)}^2/m_{\rho (770)}^2 & \lambda _3/\lambda _1 & 4.22 & 5.27 & 4.27 & 3.57 \\
\hline
 m_{a_1(1640)}^2/m_{\rho (770)}^2 & \lambda _4/\lambda _1 & 6.62 & 8.51 & 6.36 & 4.51 \\
\hline
 m_{\rho (1700)}^2/m_{\rho (770)}^2 & \lambda _5/\lambda _1 & 9.53 & 12.95 & 8.97 & 4.92 \\
\hline
\end{array}
\end{equation*}
\end{table}

\begin{table}[H]
\caption{Mass ratio calculations with $0 < \delta < 0.3$ and a maximal correction of 60\%. \\ \text{\hspace{1.5cm}} $\delta$ = 0.3000 and $\squ$ = 0.0000 minimize $\chi^2/2$ to 2.1601.}
\begin{equation*}
\begin{array}{|c|c|c|c|c|c|}
\hline
 \text{} & \lambda _n/\lambda _m & \text{Sakai-Sugimoto} & R_{n/m}^{(0)} & R_{n/m} & \text{Exp. Value }R_{n/m}^{\text{PDG}} \\
\hline
 m_{a_1(1260)}^2/m_{\rho (770)}^2 & \lambda _2/\lambda _1 & 2.32 & 2.54 & 2.39 & 2.52 \\
\hline
 m_{\rho (1450)}^2/m_{\rho (770)}^2 & \lambda _3/\lambda _1 & 4.22 & 5.27 & 4.39 & 3.57 \\
\hline
 m_{a_1(1640)}^2/m_{\rho (770)}^2 & \lambda _4/\lambda _1 & 6.62 & 8.51 & 6.57 & 4.51 \\
\hline
 m_{\rho (1700)}^2/m_{\rho (770)}^2 & \lambda _5/\lambda _1 & 9.53 & 12.95 & 9.26 & 4.92 \\
\hline
\end{array}
\end{equation*}
\end{table}

\begin{table}[H]
\caption{Mass ratio calculations with $0 < \delta < 0.4$ and a maximal correction of 60\%. \\ \text{\hspace{1.5cm}} $\delta$ = 0.4000 and $\squ$ = 0.1188 minimize $\chi^2/2$ to 1.7968.}
\begin{equation*}
\begin{array}{|c|c|c|c|c|c|}
\hline
 \text{} & \lambda _n/\lambda _m & \text{Sakai-Sugimoto} & R_{n/m}^{(0)} & R_{n/m} & \text{Exp. Value }R_{n/m}^{\text{PDG}} \\
\hline
 m_{a_1(1260)}^2/m_{\rho (770)}^2 & \lambda _2/\lambda _1 & 2.32 & 2.54 & 2.35 & 2.52 \\
\hline
 m_{\rho (1450)}^2/m_{\rho (770)}^2 & \lambda _3/\lambda _1 & 4.22 & 5.27 & 4.29 & 3.57 \\
\hline
 m_{a_1(1640)}^2/m_{\rho (770)}^2 & \lambda _4/\lambda _1 & 6.62 & 8.51 & 6.36 & 4.51 \\
\hline
 m_{\rho (1700)}^2/m_{\rho (770)}^2 & \lambda _5/\lambda _1 & 9.53 & 12.95 & 8.90 & 4.92 \\
\hline
\end{array}
\end{equation*}
\end{table}

\begin{table}[H]
\caption{Mass ratio calculations with $0 < \delta < 0.5$ and a maximal correction of 60\%. \\ \text{\hspace{1.5cm}} $\delta$ = 0.5000 and $\squ$ = 0.2238 minimize $\chi^2/2$ to 1.4707.}
\begin{equation*}
\begin{array}{|c|c|c|c|c|c|}
\hline
 \text{} & \lambda _n/\lambda _m & \text{Sakai-Sugimoto} & R_{n/m}^{(0)} & R_{n/m} & \text{Exp. Value }R_{n/m}^{\text{PDG}} \\
\hline
 m_{a_1(1260)}^2/m_{\rho (770)}^2 & \lambda _2/\lambda _1 & 2.32 & 2.54 & 2.31 & 2.52 \\
\hline
 m_{\rho (1450)}^2/m_{\rho (770)}^2 & \lambda _3/\lambda _1 & 4.22 & 5.27 & 4.18 & 3.57 \\
\hline
 m_{a_1(1640)}^2/m_{\rho (770)}^2 & \lambda _4/\lambda _1 & 6.62 & 8.51 & 6.14 & 4.51 \\
\hline
 m_{\rho (1700)}^2/m_{\rho (770)}^2 & \lambda _5/\lambda _1 & 9.53 & 12.95 & 8.54 & 4.92 \\
\hline
\end{array}
\end{equation*}
\end{table}

\begin{table}[H]
\caption{Mass ratio calculations with $0 < \delta < 0.3$ and a maximal correction of 70\%. \\ \text{\hspace{1.5cm}} $\delta$ = 0.3000 and $\squ$ = 0.0000 minimize $\chi^2/2$ to 2.1601.}
\begin{equation*}
\begin{array}{|c|c|c|c|c|c|}
\hline
 \text{} & \lambda _n/\lambda _m & \text{Sakai-Sugimoto} & R_{n/m}^{(0)} & R_{n/m} & \text{Exp. Value }R_{n/m}^{\text{PDG}} \\
\hline
 m_{a_1(1260)}^2/m_{\rho (770)}^2 & \lambda _2/\lambda _1 & 2.32 & 2.54 & 2.39 & 2.52 \\
\hline
 m_{\rho (1450)}^2/m_{\rho (770)}^2 & \lambda _3/\lambda _1 & 4.22 & 5.27 & 4.39 & 3.57 \\
\hline
 m_{a_1(1640)}^2/m_{\rho (770)}^2 & \lambda _4/\lambda _1 & 6.62 & 8.51 & 6.57 & 4.51 \\
\hline
 m_{\rho (1700)}^2/m_{\rho (770)}^2 & \lambda _5/\lambda _1 & 9.53 & 12.95 & 9.26 & 4.92 \\
\hline
\end{array}
\end{equation*}
\end{table}

\begin{table}[H]
\caption{Mass ratio calculations with $0 < \delta < 0.4$ and a maximal correction of 70\%. \\ \text{\hspace{1.5cm}} $\delta$ = 0.4000 and $\squ$ = 0.0528 minimize $\chi^2/2$ to 1.4200.}
\begin{equation*}
\begin{array}{|c|c|c|c|c|c|}
\hline
 \text{} & \lambda _n/\lambda _m & \text{Sakai-Sugimoto} & R_{n/m}^{(0)} & R_{n/m} & \text{Exp. Value }R_{n/m}^{\text{PDG}} \\
\hline
 m_{a_1(1260)}^2/m_{\rho (770)}^2 & \lambda _2/\lambda _1 & 2.32 & 2.54 & 2.34 & 2.52 \\
\hline
 m_{\rho (1450)}^2/m_{\rho (770)}^2 & \lambda _3/\lambda _1 & 4.22 & 5.27 & 4.19 & 3.57 \\
\hline
 m_{a_1(1640)}^2/m_{\rho (770)}^2 & \lambda _4/\lambda _1 & 6.62 & 8.51 & 6.14 & 4.51 \\
\hline
 m_{\rho (1700)}^2/m_{\rho (770)}^2 & \lambda _5/\lambda _1 & 9.53 & 12.95 & 8.46 & 4.92 \\
\hline
\end{array}
\end{equation*}
\end{table}

\section{Approaching the boundary of the landscape \label{landb}}

In the numerical calculations above, we impose the following restrictions:
\begin{align}
|\delta \, \lambda_n^{(1)}| < a \, \lambda_n^{(0)} \quad \forall \, n \in \{1,\ldots,5 \} \label{EQ:CorrCond}
\end{align}
where $a < 1$ is a constant. As we have only calculated the first-order corrections, a complete error analysis of the mass ratios is inaccessible. However, let us assume that condition (\ref{EQ:CorrCond}) would be applied to all orders in the $\delta$ expansion. This implies that each order term would be no higher than $a$ times the previous order term.
\begin{align}
|\delta \, \lambda_n^{(j)}| < a \, \lambda_n^{(j-1)} \quad \forall \, n \in \{1,\ldots,5 \}, \quad \forall j \in \mathbb{N}
\end{align}
With this condition, we can obtain an upper bound on the error of the eigenvalues.
\begin{align}
\Delta\lambda_n = \left|\lambda_n^{(0)} + \delta \lambda_n^{(1)} - \sum_{j=1}^\infty \delta^j \lambda_n^{(j)} \right| &\le \sum_{j=2}^\infty \left|\delta^j \lambda_n^{(j)} \right| \nonumber\\
&= \sum_{j=2}^\infty \left|\delta^{j-1}\right| \left| \delta \lambda_n^{(j)} \right| \nonumber\\
&< \sum_{j=2}^\infty \left|\delta^{j-1}\right| \left| a \lambda_n^{(j-1)} \right| \nonumber\\
&< \ldots \nonumber\\
&< \lambda^{(0)}_n \sum_{j=2}^\infty a^j \nonumber\\
&= \lambda^{(0)}_n \left( \frac{1}{1-a} - 1 - a\right) \nonumber\\
&= \lambda^{(0)}_n \left( \frac{1- (1-a)^2}{1-a}\right) \nonumber\\
&\equiv B_n(\lambda^{(0)}_n,a)
\end{align}

Then, the bound on the error of the squared-mass is simply $\Delta m_n^2 < M^2 B_n$. When we push the value of $a$ close to an uncontrollable regime ($a \sim 1$), we get better results (i.e. the $\chi^2$ improves). However, the errors also become unbounded as demonstrated in the following table:

\begin{table}[H]
\caption{Mass ratio predictions with $0 < \delta < 1$ and a maximal correction ($a$) of 99\%. \\ \text{\hspace{1.5cm}} $\delta$ = 0.9492 and $\squ$ = 0.2956 minimize $\chi^2/2$ to 0.0548.}
\begin{equation*}
\begin{array}{|c|c|c|c|c|c|c|c|c|}
\hline
 \text{} & \lambda _n/\lambda _m & \text{Sakai-Sugimoto} & R_{n/m}^{(0)} & R_{n/m} & R_{n/m}^{\text{PDG}} & \text{$\delta $ }\lambda _n^{(1)}/\lambda _n^{(0)} & \text{$\delta $ }\lambda _m^{(1)}/\lambda _m^{(0)} & B_n \\
\hline
 m_{a_1(1260)}^2/m_{\rho (770)}^2 & \lambda _2/\lambda _1 & 2.32 & 2.54 & 2.12 & 2.52 & 0.83 & 0.99 & 1468.00 \\
\hline
 m_{\rho (1450)}^2/m_{\rho (770)}^2 & \lambda _3/\lambda _1 & 4.22 & 5.27 & 3.34 & 3.57 & 0.62 & 0.99 & 3047.00 \\
\hline
 m_{a_1(1640)}^2/m_{\rho (770)}^2 & \lambda _4/\lambda _1 & 6.62 & 8.51 & 4.34 & 4.51 & 0.50 & 0.99 & 4921.00 \\
\hline
 m_{\rho (1700)}^2/m_{\rho (770)}^2 & \lambda _5/\lambda _1 & 9.53 & 12.95 & 5.22 & 4.92 & 0.39 & 0.99 & 7488.00 \\
\hline
\end{array}
\end{equation*}
\end{table}

\end{appendices}

{}

\end{document}